\documentclass[final,onefignum,onetabnum]{siamltex1213}
\usepackage{amsfonts,setspace,comment}
\usepackage{eucal}
\usepackage{amssymb,epstopdf,latexsym,comment,epsfig,subfigure}
\usepackage{graphicx,amsmath,amssymb,latexsym,psfrag,url}
\usepackage{graphics,graphicx,color}
\usepackage{enumitem}
\newcommand{\R}{\mathbb{R}}

\newcommand{\EE}{\mathbb{E}}
\newcommand{\PP}{\mathbb{P}}

\newtheorem{Theorem}{Theorem}[section]
\newtheorem{Proposition}[Theorem]{Proposition}

\newtheorem{Corollary}[Theorem]{Corollary}
\newtheorem{Lemma}[Theorem]{Lemma}

\def\ind{{\rm 1\hspace{-0.90ex}1}}

\newcommand{\tgifeps}[3]{
  \begin{figure}[t]
    \centering
    \includegraphics[width=#1cm, clip=true]{#2}
    \vspace{-0.1cm}
    \caption{#3\label{fig:#2}}
  \end{figure}

}

\title{ Modeling LRU caches with Shot Noise request processes
\thanks{A preliminary version of this work has appeared at IEEE International Conference on Computer Communication  (Infocom) 2015.}}

\author{Emilio Leonardi,
Giovanni Luca Torrisi\thanks{E . Leonardi is with  Dipartimento di Elettronica e Telecomunicazioni, Politecnico di Torino, \email{leonardi@polito.it};
G. L. Torrisi is with Istituto per le Applicazioni del Calcolo "Mauro
Picone", CNR,  \email{torrisi@iac.rm.cnr.it}. }}

\slugger{siap}{xxxx}{xx}{x}{x--x}%
\begin{document}
\maketitle


\begin{abstract}
In this paper we analyze Least Recently Used (LRU) caches operating under the Shot Noise requests Model (SNM).
The SNM was recently proposed in \cite{nostroCCR} to better capture the main characteristics of today Video on Demand (VoD) traffic.
We investigate the validity of Che's approximation \cite{che} through an asymptotic analysis
of the cache eviction time. In particular, we provide a law of large numbers, a large deviation principle and a central limit theorem for the cache eviction time,
as the cache size grows large.
Finally, we derive upper and lower bounds for the ``hit" probability in tandem networks of caches under Che's approximation.
\end{abstract}

\section{Introduction}
The design and the  analysis of caching systems, a very traditional and widely studied
topic in computer science, has recently drawn again attention
by the networking research community.
This interest revival is mainly  due to the important role that caches play today in the distribution of contents over the Internet.
Massive Content Delivery Networks, indeed,
represent the standard solution adopted by content and network providers to reach large
populations of geographically distributed users in an effective way. 
MCDN allow providers to cache contents close to the users, achieving the twofold goal of
reducing network traffic while minimizing the latency suffered by users.


Unfortunately, the performance evaluation of caching systems is very hard,
as the computational cost to analyze the behavior of a cache is exponential in both the cache size
and the number of contents~\cite{Dan1990}. For this reason, the effort of the research community has mainly focused on the
development of accurate and computationally efficient approximate techniques for the analysis of caching systems,
under various traffic conditions. Che's approximation \cite{che},
proposed for the analysis of Least Recently Used (LRU) caches under the Independent Reference Model (IRM),
has emerged as one of the most powerful methods to obtain accurate estimates of the ``hit" probability
at limited computational costs \cite{bianchi,Roberts12,nostro_infocom}. The main idea of this technique
is to summarize the response of a cache to the requests arriving for any
possible content by a single primitive quantity,
which is assumed to be deterministic and the same for any content. This approximation simplifies the analysis of
caching systems because it allows us to decouple
the dynamics of different contents. In~\cite{Roberts12} Che's approximation for LRU caches under the IRM found a theoretical justification.

Shot noise processes constitute a versatile and mathematical tractable class of stochastic models which has found several applications
in electrical engineering and queueing theory (\cite{bondesson,bordenavetorrisi,duffytorrisi,ganeshmaccitorrisi1,ganeshtorrisi,
ganeshmaccitorrisi2,lowen,mollertorrisi,stabiletorrisi,torrisi}).
In this paper we  extend the mathematical analysis of Che's approximation
to LRU caches operating under the Shot Noise Model (SNM)~\cite{nostro-tec-rep,
nostroCCR}. This model  provides a simple, flexible and accurate description of the temporal
locality found e.g. in Video on Demand (VoD) traffic, capturing today traffic characteristics in a more natural and precise way than
traditional traffic models. Inspired by the seminal paper \cite{Roberts12}, we investigate the validity of Che's approximation by means of an asymptotic analysis
of the cache eviction time. Specifically, we provide a law of large numbers, large deviations and
a central limit theorem for the cache eviction time, as the cache size grows large. Furthermore,
to the best of our knowledge, we give for the first time a non-asymptotic analytical
upper bound for the error estimate of the ``hit" probability entailed by Che's approximation.
We provide also upper and lower bounds for the ''hit" probability, under
Che's approximation, for a tandem network of caches. Finally, we present some numerical illustrations.
Our results show that Che's approximation is a
provable, highly accurate and scalable tool to assess the performance of LRU caching systems under the SNM.

\section{System description and motivations}

We consider a cache, whose size (or capacity), expressed
in number of objects (or contents), is denoted by $C$.
The cache is fed by an exogenous arrival process of objects' requests
generated by users. Requests which find the object in the cache
are said to produce a ``hit", whereas requests that do not find the object in
the cache are said to produce a "miss". An important performance index
is the ``hit" probability, which is the fraction of the requests producing a ``hit".
The miss stream of a cache, i.e. the
process of requests which are not locally satisfied by the cache,
is forwarded to either other caches or to a common repository containing all the objects, i.e. the entire objects' catalogue.
In the literature it is common to neglect all propagation delays.

In this paper we focus on caches implementing the LRU policy:
upon the arrival of a request, an object not already stored in the cache is inserted into it.
If the cache is full, to make room for a new object the least recently used item is evicted, i.e. the
object which has not been requested for the longest time is expunged from the cache.

Several models have been proposed to describe the process of requests arriving at a cache.
The simplest and still the most widely adopted is certainly the IRM
~\cite{Coffman:73}, which  makes the following two fundamental assumptions: $i)$ the catalogue consists of a fixed number of objects,
which does not change over the time; $ii)$ the process of requests of a given object is modeled by a homogeneous Poisson process.
As a consequence, the IRM completely ignores all temporal correlations
in the sequence of requests and does not take into account
a key feature of real traffic referred to as temporal locality,
which means that if an object is requested at a given time,
then it is more likely that the same object will be requested again in the near future.
It is well-known that the temporal locality has a beneficial effect on the cache performance, as
it increases the ``hit" probability ~\cite{Coffman:73}.

Several extensions of the IRM have been proposed to incorporate the temporal locality into the traffic model.
Existing generalizations \cite{bianchi,Coffman:73,virtamo98}
typically assume that the process of requests is
time-stationary, usually either a renewal process or a Markov modulated
Poisson process. However, these models do not capture the kind of temporal locality encountered
in traces related to Video on Demand (VoD) traffic, which is
instead well described by the SNM  as shown in \cite{nostro-tec-rep,nostroCCR}.

\section{Shot Noise Model and Cache Analysis}\label{sec:traffic}

The basic idea of the SNM is to represent the requests' process as the superposition of many independent processes,
each one referring to a specific object. The requests' process of a fixed content $m$ is specified by two physical (random) parameters:
$\xi_m$ and $Z_m$. $\xi_m$ represents the time instant at which the content enters the system (i.e. it becomes available to the users); mark $Z_m$ describes some attribute of the content $m$,
which summarizes its main characteristics (content type, volume, etc.).

We assume that the set of times $N\equiv\{\xi_m\}_{m\geq 1}$ at which contents become available to users  (i.e., they are introduced in the common repository)
is distributed according to a homogeneous Poisson process on $\R$
with intensity $\lambda>0$. Here, $\{\xi_m\}_{m\geq 1}$ is supposed to be an unordered set of times.
We suppose that, after the introduction into the catalogue of the content $m$, the requests for this content arrive at the cache according to a Cox process
$N^{(m)}\equiv\{T_n^{(m)}\}_{n\geq 1}$ on $\R$ whose stochastic
intensity $\{\lambda_m(t)\}_{t\in\R}$ is defined by
\[
\lambda_m(t):=h(t-\xi_m,Z_m),
\]
see e.g. \cite{daley}.
We assume that $\{Z_m\}_{m\geq 1}$ is a sequence of independent and identically distributed random variables,
independent of $\{\xi_m\}_{m\geq 1}$, with values on some measurable
space $(E,\mathcal E)$. Furthermore, we suppose that $h:\R\times E\to [0,\infty)$ is a
measurable non-negative function such that $h(t,z)=0$ for any $t<0$ and $z\in
E$.
Finally, we suppose that, for any $m\geq 1$, $T_1^{(m)}<T_2^{(m)}<\ldots$ almost surely
and we assume that the Cox processes $\{N^{(m)}\}_{m\geq 1}$ are independent,
given $\{(\xi_m,Z_m)\}_{m\geq 1}$.

\subsection{Formal definition of the cache eviction time}\label{subsec:def}

We denote by $m_0$ a tagged content introduced into the catalogue at the deterministic
time $\xi_{m_0}=-x$, $x>0$, and requested at time $0$. Moreover, we denote by
$\bold{X}_{m_0}(t)$, $t>0$, the number of contents different from $m_0$ that
have been requested in the time interval $[0,t]$, i.e.
\begin{align}
\bold{X}_{m_0}(t)=\sum_{m\neq m_0}\ind\{m \mbox{ requested in } [0,t],\,\xi_m\in (-\infty,t]\}.\nonumber
\end{align}
Throughout this paper we shall consider the random variable
\[
X_{m_0,x}(t):=\bold{X}_{m_0}(t)\,|\,\xi_{m_0}=-x,\quad t,x>0
\]
which plays an important role in the dynamics of an LRU cache because the cache eviction time may be expressed in terms of $X_{m_0,x}(t)$.
Indeed, under the LRU replacement policy, 
we have that the content $m_0$ is expunged from the cache (provided it is not requested again after time 0) as soon as the $C$th content, different from
$m_0$, is requested.
So, under
the LRU replacement policy, the so-called cache eviction time for the content $m_0$ is given by the random variable
\[
T_C(m_0,x):=\inf\{t>0:\,\,X_{m_0,x}(t)=C\},
\]
where once again we remark that, by construction, $T_C(m_0,x)$ is the time at which the content $m_0$
is expunged from the cache, provided that no requests for the content $m_0$ are observed in the time interval $(0,T_C(m_0,x)]$.

\subsection{The distribution of $X_{m_0,x}(t)$}\label{subsec:poisson}

Define the quantity
\begin{equation}\label{eq:grev}
g(t):=\int_{0}^\infty\mathbb{E}\left[1-
\mathrm{e}^{-\int_{u-t}^{u}h(s,Z_1)\,\mathrm{d}s}\right]\,\mathrm{d}u,\quad t>0.
\end{equation}
The following proposition holds.
\begin{Proposition}\label{prop:lap}
If $g(t)<\infty$, then the random variable $X_{m_0,x}(t)$ is Poisson distributed with mean $\lambda g(t)$.
\end{Proposition}

Note that the condition $g(t)<\infty$ is fairly general: for example, it is satisfied whenever the popularity
profile is of multiplicative form i.e. (with a little abuse of notation)
\begin{equation}\label{eq:popprof}
h(t,z):=zh(t),\quad t\in \R,\,z\in E\subseteq (a,\infty),\,a>0
\end{equation}
and
\begin{equation}\label{eq:hintthintmeanz}
\text{$h\equiv 0$ on $(-\infty,0)$, $\int_0^\infty h(t)\,\mathrm{d}t=1$ and $\mathbb{E}[Z_1]<\infty$.}
\end{equation}
Indeed, in such a case, for $t>0$, we have
\begin{align}
g(t)&=\int_{0}^\infty\left[1-\phi_{Z_1}\left(-\int_{u-t}^{u}h(s)\,\mathrm{d}s\right) \right]\,\mathrm{d}u \label{eqg}\\
&\leq\EE[Z_1]\int_0^\infty\left(\int_{u-t}^{u}h(s)\,\mathrm{d}s\right)\,\mathrm{d}u\leq\EE[Z_1]t
<\infty,\nonumber
\end{align}
where $\phi_{Z_1}(\theta):=\EE[\exp(\theta Z_1)]$, $\theta\in\R$, and we used the elementary inequality $\mathrm{e}^x\geq 1+x$, $x\in\R$.

\noindent{\it Proof\,\,of\,\,Proposition\,\,\ref{prop:lap}.}
For any $t_0>t>0$, we define the
''restriction'' of $\bold{X}_{m_0}(t)$ to contents that have been introduced in
the model in the time interval $[t- t_0,t]$ by
\begin{align}
\bold{X}_{m_0}^{(t_0)}(t):=\sum_{m\neq m_0}\ind\{m \mbox{ requested in } [0, t ],\,\xi_m\in [t-t_0,t]\}.
\nonumber
\end{align}
By the Slivnyak-Mecke theorem (see e.g. Proposition 13.1.VII p. 281 in \cite{daley2}), the law of
$\{\xi_m\}_{m\neq m_0}$ given the event
$\{\xi_{m_0}=-x\}$ coincides with the law of
$\{\xi_m\}_{m\geq 1}$ and so, for any $\theta\in\R$,
\begin{equation}\label{eq:1}
\mathbb{E}\left[\mathrm{e}^{\theta
\bold{X}_{m_0}^{(t_0)}(t)}\,\mid\,\xi_{m_0}=-x\right]
=\mathbb{E}\left[\mathrm{e}^{\theta
\widetilde{\bold{X}}_{t_0}(t)}\right],
\end{equation}
where
\[
\widetilde{\bold{X}}_{t_0}(t):=\sum_{m\geq 1}\ind\{m \mbox{ requested
in  } [0, t ],\,\xi_m\in [t-t_0,t]\}.
\]
Letting $N([t-t_0,t])$ denote the number of points
$\{\xi_m\}_{m\geq 1}$ in the time interval $[t-t_0,t]$ and
$N^{(m)}([0,t])$ denote the number of points
$\{T_n^{(m)}\}_{n\geq 1}$ in the time interval $[0,t]$, we
rewrite $\widetilde{\bold{X}}_{t_0}(t)$ as
\[
\widetilde{\bold{X}}_{t_0}(t)=\sum_{m=
1}^{N([t-t_0,t])}\ind\{N^{(m)}([0,t])\geq 1\}.
\]
Since, given $\xi_m$ and $Z_m$, $N^{(m)}$ is a Poisson process with
intensity function $h(\cdot-\xi_m,Z_m)$, we have
\begin{align}
p_t(\xi_m,Z_m):=\PP(N^{(m)}([0,t])\geq 1\,|\,\xi_m,Z_m)=
1-\mathrm{e}^{-\int_{0}^{t}h(s-\xi_m,Z_m)\,\mathrm{d}s}.\nonumber
\end{align}
Recalling that, given $\{N([t-t_0,t])=k\}$, the $k$ points of $N$ on $[t-t_0,t]$
are independent and uniformly distributed over $[t-t_0,t]$ (see e.g. \cite{daley}),
for any $\theta\in\R$, we have
\begin{align}
\mathbb{E}\left[\mathrm{e}^{\theta\widetilde{\bold{X}}_{t_0}(t)}\,\mid\,
N([t-t_0,t])=k\right]&=\mathbb{E}\left[\prod_{m=1}^{k}\mathrm{e}^{\theta\ind\{N^{(m)}([0,t])\geq
1\}}\,\mid\, N([t-t_0,t])=k\right]\nonumber\\
&=\left(1+(\mathrm{e}^\theta-1)\frac{1}{t_0}\int_{t-t_0}^t\mathbb{E}
[p_t(u,Z_1)]\,\mathrm{d}u\right)^k.\nonumber
\end{align}
Therefore, since $N([t-t_0,t])$ is Poisson distributed with mean $\lambda t_0$, we have
\begin{align}
\mathbb{E}\left[\mathrm{e}^{\theta\widetilde{\bold{X}}_{t_0}(t)}\right]&
=\exp\left(\lambda(\mathrm{e}^\theta-1)\int_{-t_0}^0\mathbb{E}
\left[p_t(u+t,Z_1)\right]\,\mathrm{d}u\right).\label{eq:2}
\end{align}
The claim follows by \eqref{eq:1} and \eqref{eq:2}, letting $t_0$
tend to $\infty$.
\\
\noindent$\square$

In the context of an LRU cache under the SNM, Che's approximation consists in
replacing the cache eviction time $T_C(m_0,x)$ by the deterministic constant
\[
t_C(m_0,x):=\inf\{t>0:\,\,\mathbb{E}[X_{m_0,x}(t)]=C\}.
\]
Note that, if $g(t)<\infty$ for any $t>0$, then by Proposition \ref{prop:lap} we have
\[
t_C(m_0,x)=\inf\{t>0:\,\,\lambda g(t)=C\}.
\]
So, if moreover $g:(0,\infty)\to (0,\infty)$ is strictly increasing,
we deduce
\begin{equation}\label{eq:tc}
t_C(m_0,x)=g^{-1}(C/\lambda).
\end{equation}
Since the law of $X_{m_0,x}(t)$ (and therefore of $T_C(m_0,x)$) does not depend on $m_0$ and $x$,
hereafter we simply write $X(t)$, $T_C$ and $t_C$
in place of $X_{m_0,x}(t)$, $T_C(m_0,x)$ and $t_C(m_0,x)$.

\subsection{Asymptotic analysis of $T_C$}

In this subsection we investigate the validity of Che's approximation for large values of $C$. We shall do this
by analyzing the behavior of $T_C$ as $C\uparrow\infty$.
Intuitively, Che's approximation finds a theoretical justification if we may show that, as
$C\uparrow\infty$, $T_C/t_C \to 1$ almost surely.
This is indeed achieved in Proposition \ref{cor:lln}. Proposition \ref{cor:ldp} provides asymptotic tail estimates for $T_C$ and Corollary \ref{cor:rem1}
gives asymptotic upper and lower bounds for the probability that $T_C$ deviates from its most probable value $t_C$, as $C$ grows large.
Finally, the Gaussian approximation for $T_C$ in Proposition \ref{prop:normalcaching} allows us to construct asymptotic
confidence intervals for $T_C$,
see the short discussion after the statement of Proposition \ref{prop:normalcaching}.
Hereafter, we shall consider the function $g$ defined by \eqref{eq:grev}.

\subsubsection{Law of large numbers and tail estimates for the cache eviction time}\label{subsec:ldp}

The following law of large numbers and tail estimates hold.
\begin{Proposition}\label{cor:lln}
If $g:(0,\infty)\to (0,\infty)$ is strictly increasing, $g,g^{-1}:(0,\infty)\to (0,\infty)$ are bijective and continuous
(i.e. $g$ is a homeomorphism of $(0,\infty)$) and,
for any divergent sequences $\{a_n\}_{n\geq 1}$, $\{b_n\}_{n\geq 1}$ of positive numbers,
\begin{equation}\label{eq:glim}
\lim_{n\to\infty}g(a_n)/g(b_n)=1\Rightarrow\lim_{n\to\infty}a_n/b_n=1,
\end{equation}
then
\begin{equation}\label{eq:cheas}
\lim_{C\to\infty}\frac{T_C}{t_C}=1,\quad\text{almost surely.}
\end{equation}
\end{Proposition}
\begin{Proposition}\label{cor:ldp}
If $g:(0,\infty)\to (0,\infty)$ is strictly increasing and $g,g^{-1}:(0,\infty)\to (0,\infty)$ are bijective and continuous,
then
\begin{align}
\lim_{C\to\infty}&\frac{1}{C}\log\PP(T_C>g^{-1}(Cx_r))=-I(x_r),\quad\text{$\forall$ $x_r>1/\lambda$}\label{eq:tail}
\end{align}
and
\begin{align}
\lim_{C\to\infty} & \frac{1}{C}\log\PP(T_C\leq g^{-1}(Cx_l))=-I(x_l),\quad\text{$\forall$ $x_l\in (0,1/\lambda)$,}\label{eq:tail1}
\end{align}
where
\begin{equation}\label{eq:rfI}
I(x):=\lambda x-1-\log(\lambda x),\quad x>0\quad\text{and}\quad I(0):=+\infty.
\end{equation}
\end{Proposition}
\begin{Corollary}\label{cor:rem1}
Under the assumptions of Proposition \ref{cor:ldp}, we have that
for any $\delta\in (0,1)$ and $\varepsilon>0$ there exists $C_{\delta,\varepsilon}$ so that
for any $C>C_{\delta,\varepsilon}$
\begin{align}
\mathrm{e}^{-C(I(g(t_C(1+\delta))/C)+\varepsilon)}&\leq\PP(T_C>t_C(1+\delta))\leq\mathrm{e}^{-C(I(g(t_C(1+\delta))/C)-\varepsilon)}\label{eq:LD1rev}
\end{align}
and
\begin{align}
\mathrm{e}^{-C(I(g(t_C(1-\delta))/C)+\varepsilon)}&\leq\PP(T_C\leq t_C(1-\delta))\leq\mathrm{e}^{-C(I(g(t_C(1-\delta))/C)-\varepsilon)},\label{eq:LD2rev}
\end{align}
where the function $I$ is defined by \eqref{eq:rfI}.
\end{Corollary}

The proofs of Propositions \ref{cor:lln} and \ref{cor:ldp} are based on a large deviation principle for the process $\{g(T_C)/C\}_{C\geq 1}$
stated in Lemma \ref{prop:ldp}. We recall, see e.g. \cite{dembo}, that a non-negative stochastic process $\{Y(t)\}_{t\geq 0}$ obeys a large deviation principle on $[0,\infty)$ with speed
$v$ and rate function $J$ if $v:[0,\infty)\to [0,\infty)$ is a function which increases to infinity and $J:[0,\infty)\to [0,\infty]$
is a lower semi-continuous function such that, for all Borel sets $B\subset [0,\infty)$,
\begin{align}
-\inf_{x\in B^\circ }J(x) \leq\liminf _{t\to\infty}\frac{1}{v(t)}\log\PP(Y(t)\in B)
\leq\limsup_{t\to\infty}\frac{1}{v(t)}\log \PP(Y(t)\in B)\leq-\inf_{x\in \overline{B}}J(x),\nonumber
\end{align}
where $B^\circ$ denotes the interior of $B$ and $\overline{B}$ denotes the closure of $B$.

For later purposes, we recall
that a rate function $J$ on $[0,\infty)$ has no peaks if $(i)$ there exists $\bar x\in (0,\infty)$ such that $J(\bar x)=0$; $(ii)$
$J$ is non-increasing on $(0,\bar x)$ and non-decreasing on $(\bar x,\infty)$.

\begin{Lemma}\label{prop:ldp}
Under the assumptions of Proposition \ref{cor:ldp}, we have that the family of random variables $\{g(T_C)/C\}_{C\geq 1}$
obeys a large deviation principle on $[0,\infty)$ with speed $v(C):=C$ and rate function $J:=I$ defined by \eqref{eq:rfI}.
\end{Lemma}

\noindent{\bf Remark\,\,1.}
For later purposes, we remark that the rate function $I$ defined in \eqref{eq:rfI} is continuous on $(0,\infty)$,
$I(1/\lambda)=0$ and $I$ decreases on $(0,1/\lambda)$ and increases on $(1/\lambda,\infty)$. So, in particular, $I$ has no peaks.
\\
\noindent$\square$

\noindent{\bf Remark\,\,2.}
Here, we assume that $g$ defined by \eqref{eq:grev} is a strictly increasing homeomorphism of $(0,\infty)$,
and we give sufficient conditions which guarantee \eqref{eq:glim}.\\
\noindent$(i)$ If $g^{-1}$ is ultimately Lipschitz continuous, i.e.
\[
\text{there exist $K_1,K_2>0$ such that $|g^{-1}(x)-g^{-1}(y)|\leq K_1|x-y|$, for any $x,y>K_2$}
\]
then \eqref{eq:glim} holds. Indeed, let $\varepsilon>0$ be arbitrarily fixed and $n_\varepsilon^{(1)}\geq 1$ so large that
$g(b_n)>K_2+\varepsilon$ for all $n\geq n_\varepsilon^{(1)}$. By the Lipschitz property of $g^{-1}$ we have
\begin{align}
\sup_{n\geq n_\varepsilon^{(1)}}|g^{-1}(g(b_n)\pm\varepsilon)-b_n|&=\sup_{n\geq n_\varepsilon^{(1)}}|g^{-1}(g(b_n)\pm\varepsilon)-g^{-1}(g(b_n))|\leq K_1\varepsilon.\nonumber
\end{align}
Therefore
\begin{equation}\label{eq:dis1}
b_n-K_1\varepsilon<g^{-1}(g(b_n)-\varepsilon)\quad\text{and}\quad g^{-1}(g(b_n)+\varepsilon)<b_n+K_1\varepsilon,\quad\text{$\forall$ $n\geq n_\varepsilon^{(1)}$.}
\end{equation}
By assumption $g(a_n)/g(b_n)\to 1$, as $n\to\infty$. Therefore there exists $n_\varepsilon^{(2)}\geq 1$ such that for any $n\geq n_\varepsilon^{(2)}$
\begin{equation}\label{eq:dis2}
g^{-1}(g(b_n)-\varepsilon)<a_n<g^{-1}(g(b_n)+\varepsilon).
\end{equation}
The claim follows combining the inequalities \eqref{eq:dis1} and \eqref{eq:dis2}.
\\
\noindent$(ii)$ If there exists $\bar t>0$ such that $g$ is differentiable on $(\bar t,\infty)$ and $\inf_{t>\bar t}g'(t)>0$ then \eqref{eq:glim} holds.
Indeed, for all $x>g(\bar t)$ we have
\[
0<(g^{-1})'(x)=1/g'(g^{-1}(x))\leq (\inf_{t>\bar t}g'(t))^{-1},
\]
therefore $(g^{-1})'$ is ultimately bounded and the claim follows by the previous point $(i)$.\\
\noindent$\square$

\noindent{\bf Example\,\,1.}
Consider the SNM defined by a multiplicative popularity profile of the form
\eqref{eq:popprof} and assume \eqref{eq:hintthintmeanz}.
In such a case, $g$ is given by \eqref{eqg}
and it clearly satisfies the assumptions of Proposition \ref{cor:ldp}. We check that $g$ satisfies \eqref{eq:glim}.
Setting $H(t):=\int_0^t h(s)\,\mathrm{d}s$, $t>0$, we have
\begin{align}
g(t) &=\int_{0}^\infty\left[1- \phi_{Z_1}\left(-H(u)+H(u-t)\right) \right]\,\mathrm{d}u \nonumber\\
    & =  \int_{0}^t\left[1- \phi_{Z_1}\left(-H(u)\right) \right]\,\mathrm{d}u + \int_{0}^\infty\left[1- \phi_{Z_1}\left(-H(u+t)
    +H(u)\right) \right]\,\mathrm{d}u.\label{eq:secondbounded}
\end{align}
Since  $\phi_{Z_1}(\cdot)$ is differentiable on $(-\infty,0)$, we easily have that $g(\cdot)$ is differentiable on $(0,\infty)$ and
\begin{align}
g'(t)&= 1- \phi_{Z_1}\left(-H(t)\right)+\int_{0}^\infty \phi'_{Z_1}\left(-H(u+t)+H(u)\right)h(u+t)\,\mathrm{d}u\label{eq:deriv}\\
&\geq 1- \phi_{Z_1}\left(-H(t)\right),\nonumber
\end{align}
where the latter inequality follows by the nonnegativity of the third addend in the right-hand-side of \eqref{eq:deriv}.
The claim follows by Remark 2$(ii)$ noticing that if $\bar{t}$ is such that $H(\bar{t})>0$, then
\begin{equation*}
\inf_{t>\bar{t}}g'(t)\geq 1-\phi_{Z_1}(-H(\bar{t}))>0.
\end{equation*}

In the particular case,  when
\begin{equation}\label{eq:h}
h(t,z):=\frac{z}{L}\ind_{[0,L]}(t),\quad\text{for some constant $L>0$}
\end{equation}
we have
\begin{align}
g(t)&=2(t\wedge L)+(t\vee L-t\wedge L)\left(1-\phi_{Z_1}\left(-(t\wedge L)/L\right)\right)
-2\mathbb{E}\left[\frac{L}{Z_1}\left(1-\mathrm{e}^{-\frac{(t\wedge L)Z_1}{L}}\right)\right],\label{eq:g1}
\end{align}
where for $a,b\in\mathbb R$ we set $a\wedge b:=\min\{a,b\}$ and $a\vee b:=\max\{a,b\}$. Indeed, for $t>0$, we have
\begin{align}
g(t)&=\int_{0}^\infty\Big(1-\phi_{Z_1}\Big(-\frac{1}{L}\int_{u-t}^{u}\ind_{[0,L]}(s)\,\mathrm{d}s\Big)\Big)\,\mathrm{d}u
=\int_{0}^\infty\Big(1-\phi_{Z_1}(-\eta(t,u))\Big)\,\mathrm{d}u,\nonumber
\end{align}
where
\begin{align}
\eta(t,u):=\frac{1}{L}\int_{u-t}^{u}\ind_{[0,L]}(s)\,\mathrm{d}s=\frac{(u\wedge L-(u-t)^+)^+}{L}\nonumber
\end{align}
and for $a\in\mathbb R$ we set $a^+:=a\vee 0$.
We distinguish two cases: $0<t\leq L$ and $t>L$. If $0<t\leq L$, then if $u<t$ then $u<L$. So, for $t\in (0,L]$,
\begin{align}
g(t)&=\int_{0}^{t}\Big(1-\phi_{Z_1}\Big(-\frac{u}{L}\Big)\Big)\,\mathrm{d}u
+\int_{t}^{L+t}\Big(1-\phi_{Z_1}\Big(-\frac{1}{L}(t\ind_{(0,L]}(u)+(t-u+L)\ind_{(L,L+t]}(u)\Big)\Big)\,\mathrm{d}u
\nonumber\\
&=t-\mathbb{E}\left[\frac{L}{Z_1}\left(1-\mathrm{e}^{-\frac{Z_1}{L}t}\right)\right]+(L-t)\Big(1-\phi_{Z_1}\Big(-\frac{t}{L}\Big)\Big)+
\int_{L}^{L+t}\Big(1-\phi_{Z_1}\Big(-\frac{1}{L}(t-u+L)\Big)\Big)\,\mathrm{d}u\nonumber\\
&=2t+(L-t)\Big(1-\phi_{Z_1}\Big(-\frac{t}{L}\Big)\Big)-2\mathbb{E}\left[\frac{L}{Z_1}\left(1-\mathrm{e}^{-\frac{Z_1}{L}t}\right)\right].\nonumber
\end{align}
If $t>L$, then if $u\geq t$ then $u\geq t>L$. So, for $t>L$,
\begin{align}
g(t)&
=\int_{0}^{L}\Big(1-\phi_{Z_1}\Big(-\frac{u}{L}\Big)\Big)\,\mathrm{d}u+
(t-L)(1-\phi_{Z_1}(-1))
+\int_{t}^{L+t}\Big(1-\phi_{Z_1}\Big(-\frac{t-u+L}{L}\Big)\Big)\,\mathrm{d}u\nonumber\\
&=L-\EE\left[\frac{L}{Z_1}\left(1-\mathrm{e}^{-Z_1}\right)\right]
+(t-L)(1-\phi_{Z_1}(-1))+L-\EE\left[\frac{L}{Z_1}\left(1-\mathrm{e}^{-Z_1}\right)\right]\nonumber\\
&=2L
+(t-L)(1-\phi_{Z_1}(-1))-2\mathbb{E}\left[\frac{L}{Z_1}\left(1-\mathrm{e}^{-Z_1}\right)\right].\nonumber
\end{align}
\\
\noindent$\square$

\noindent${\it Proof\,\,of\,\, Proposition\,\,\ref{cor:lln}.}$ It is well-known that
\begin{equation}\label{eq:lln}
\lim_{C\to\infty}\frac{g(T_C)}{C}=1/\lambda,\quad\text{almost surely}
\end{equation}
if and only if
\[
\PP\left(\bigcap_{n\geq 1}\bigcup_{C\geq n}\Big|\frac{g(T_C)}{C}-\frac{1}{\lambda}\Big|>\varepsilon\right)=0,\quad\text{for any $\varepsilon>0$.}
\]
Therefore \eqref{eq:lln} follows by the Borel-Cantelli lemma if we check that
\[
\sum_{C\geq 1}\PP\left(\Big|\frac{g(T_C)}{C}-\frac{1}{\lambda}\Big|>\varepsilon\right)<\infty,\quad\text{for any $\varepsilon>0$.}
\]
Let $\varepsilon\in (0,\lambda^{-1})$ be arbitrarily fixed and take $\delta\in (0,I(\lambda^{-1}-\varepsilon)\wedge I(\lambda^{-1}+\varepsilon))$. By Lemma \ref{prop:ldp} we have
that there exists a non-negative integer $C_\delta$ such that for any $C\geq C_\delta$
\begin{align}
\PP\left(\frac{g(T_C)}{C}\geq\lambda^{-1}+\varepsilon\right)&\leq\mathrm{e}^{-\left(\inf_{x\geq \lambda^{-1}+\varepsilon}I(x)-\delta\right)C}=\mathrm{e}^{-(I(\lambda^{-1}+\varepsilon)-\delta)C}\nonumber
\end{align}
and
\begin{align}
\PP\left(\frac{g(T_C)}{C}\leq\lambda^{-1}-\varepsilon\right)&\leq\mathrm{e}^{-\left(\inf_{x\leq \lambda^{-1}-\varepsilon}I(x)-\delta\right)C}=\mathrm{e}^{-(I(\lambda^{-1}-\varepsilon)-\delta)C},\nonumber
\end{align}
where we used Remark 1. Therefore
\begin{align}
\sum_{C\geq C_\delta}\PP\left(\Big|\frac{g(T_C)}{C}-\frac{1}{\lambda}\Big|>\varepsilon\right)&\leq
\sum_{C\geq C_\delta}\PP\left(\frac{g(T_C)}{C}\geq\lambda^{-1}+\varepsilon\right)+\sum_{C\geq C_\delta}\PP\left(\frac{g(T_C)}{C}\leq\lambda^{-1}-\varepsilon\right)\nonumber\\
&\leq\sum_{C\geq C_\delta}\mathrm{e}^{-(I(\lambda^{-1}+\varepsilon)-\delta)C}+\sum_{C\geq C_\delta}\mathrm{e}^{-(I(\lambda^{-1}-\varepsilon)-\delta)C}<\infty,\nonumber
\end{align}
which proves \eqref{eq:lln}. Then the claim follows by assumption \eqref{eq:glim} noticing that by \eqref{eq:lln} and relation $t_C=g^{-1}(C/\lambda)$ we easily get
$g(T_C)/g(t_C)\to 1$ almost surely, as $C\to\infty$.
\\
\noindent$\square$

\noindent${\it Proof\,\,of\,\,Prooposition\,\,\ref{cor:ldp}.}$ By Remark 1 we have that,
for any $x_r>1/\lambda$, $\inf_{y>x_r}I(y)=\inf_{y\geq x_r}I(y)=I(x_r)$,
and, for any $x_l\in (0,1/\lambda)$, $\inf_{y<x_l}I(y)=\inf_{y\leq x_l}I(y)=I(x_l)$.
Relations \eqref{eq:tail} and \eqref{eq:tail1} follow by applying the large deviation principle of Lemma \ref{prop:ldp}
considering, respectively, the Borel sets $B=(x_r,\infty)$ and $B=(0,x_l)$ (note that $g^{-1}$ is strictly increasing since $g$
is such).
\\
\noindent$\square$

\noindent${\it Proof\,\,of\,\,Corollary\,\,\ref{cor:rem1}.}$ The claim easily follows by \eqref{eq:tc}, \eqref{eq:tail} and \eqref{eq:tail1}.
\\
\noindent$\square$

The proof of Lemma \ref{prop:ldp} uses a result from \cite{duffield}, which we recall for the sake of clarity.
We first introduce some notation and terminology. Let $\{Y(t)\}_{t\geq 0}$ be a non-negative stochastic process whose sample paths are
right-continuous, non-decreasing and such that $\lim_{t\to\infty}Y(t)=\infty$ almost surely. We define the inverse process of
$\{Y(t)\}_{t\geq 0}$ as
\[
W(z):=\inf\{t\geq 0:\,\,Y(t)\geq z\},\quad z\geq 0.
\]
The following theorem holds (see Theorem 1$(i)$ in \cite{duffield}).

\begin{Theorem}\label{thm:duffield}
Let $\{Y(t)\}_{t\geq 0}$ and $\{W(z)\}_{z\geq 0}$ be as above and
let $v:(0,\infty)\to (0,\infty)$ be a strictly increasing homeomorphism of $(0,\infty)$.
We have that if $\{Y(t)/v(t)\}_{t\geq 0}$ obeys a large deviation principle on $[0,\infty)$ with speed $v$
and rate function $I$ which has no peaks, then $\{v(W(z))/z\}_{z>0}$ obeys a large deviation principle on $[0,\infty)$ with speed $\tilde{v}(z):=z$ and rate function
$\tilde{I}(z):=zI(1/z)$, $z>0$, $\tilde{I}(0):=\lim_{z\to 0^+}zI(1/z)$.
\end{Theorem}

\noindent${\it Proof\,\,of\,\,Lemma\,\,\ref{prop:ldp}.}$ Let $\theta\in\mathbb R$ be arbitrarily fixed.
By Proposition \ref{prop:lap} $X(t)$ is Poisson distributed with mean $\lambda g(t)$. Therefore
\begin{align}
\lim_{t\to\infty}\frac{1}{g(t)}\log\EE\left[\mathrm{e}^{\theta X(t)}\right]&=
\lim_{t\to\infty}\frac{1}{g(t)}\log\mathrm{e}^{\lambda g(t)(\mathrm{e}^\theta-1)}=\lambda(\mathrm{e}^\theta-1):=\Lambda(\theta).\nonumber
\end{align}
So by the G\"artner-Ellis theorem (see e.g. \cite{dembo})
the stochastic process $\{X(t)/g(t)\}_{t\geq 1}$ satisfies a large deviation principle on $[0,\infty)$
with speed $g$ and rate function $\Lambda^*(x):=\sup_{\theta\in\mathbb R}(\theta x-\Lambda(\theta))
=\lambda-x+x\log(x/\lambda)$, $x>0$, $\Lambda^*(0):=\lambda$.
Note that $\{T_C\}_{C\geq 1}$ is the inverse process of $\{X(t)\}_{t\geq 0}$.
The claim then follows by Theorem \ref{thm:duffield}.
Indeed the rate function $\Lambda^*$ has no peaks since $\Lambda^*(\lambda)=0$ and $\Lambda^*$ decreases on $(0,\lambda)$
and increases on $(\lambda,\infty)$.
\\
\noindent$\square$

\subsubsection{Normal approximation of the cache eviction time}\label{subsec:gauss}

Hereafter, we denote by $\mathcal{N}(0,1)$ a standard normal random variable
and by $\overset{law}\longrightarrow$ the convergence in distribution. Following the ideas in \cite{Roberts12} (see Propositions 1 and 3 therein), we derive a
central limit theorem for the cache eviction time of the SNM.
\begin{Proposition}\label{prop:normalcaching}
Assume $g:(0,\infty)\to (0,\infty)$ bijective, strictly increasing and such that there exists a positive function $f$ such that
\begin{align}
\lim_{y\to\infty}f(y)\in [0,\infty]\quad\text{and}\quad
\lim_{y\to\infty}\frac{g(y)-g(y+xf(y))}{\sqrt{g(y+xf(y))}}=-\frac{x}{\sqrt\lambda}.
\label{eq:HYP1}
\end{align}
Then
\begin{equation}\label{varsmall}
\frac{T_C-t_C}{f(t_C)}\overset{law}\longrightarrow\mathcal{N}(0,1),\quad\text{as $C\to\infty$.}
\end{equation}
\end{Proposition}
Under the assumptions of Proposition \ref{prop:normalcaching} one can construct asymptotic confidence intervals for $T_C$.
Indeed, if $\nu>0$ is such that $\mathbb P(|\mathcal{N}(0,1)|\leq\nu)=\mu\in (0,1)$, then, as $C\to\infty$,
$[t_C-\nu f(t_C),t_C+\nu f(t_C)]$ is an asymptotic confidence interval for $T_C$ at the level $\mu$, as the following simple computation shows:
\begin{align}
\mathbb{P}\left(t_C-\nu f(t_C)\leq T_C\leq t_C+\nu f(t_C)\right)&=\mathbb{P}\left(\Big|\frac{T_C-t_C}{f(t_C)}\Big|\leq\nu\right)\nonumber\\
&\simeq\mathbb{P}(|\mathcal{N}(0,1)|\leq\nu)=\mu,\quad\text{as $C\to\infty$}.\nonumber
\end{align}
Clearly, for a fixed level $\mu$, by using the tables of the Gaussian distribution one finds the value $\nu$ which determines
the asymptotic confidence interval.

\noindent{\bf Example\,\,2.}
Consider the SNM defined by a multiplicative popularity profile of the form
\eqref{eq:popprof} and assume \eqref{eq:hintthintmeanz} and
\[
\int_0^\infty t h(t)\,\mathrm{d}t<\infty.
\]
In such a case, $g$ is given by \eqref{eqg}
and, as noticed in Example 1, $g$ is an increasing homeomorphism of $(0,\infty)$.
We shall check later on that \eqref{eq:HYP1} holds with
\begin{equation}\label{eq:fx}
f(x):=\sqrt{\frac{x}{\lambda(1-\phi_{Z_1}(-1))}}.
\end{equation}
Therefore we have the normal approximation \eqref{varsmall} and
asymptotic confidence intervals for $T_C$ can be constructed as described above.
To verify \eqref{eq:HYP1} we start noticing that
\begin{equation}\label{eq:limg}
\lim_{t\to\infty}\frac{g(t)}{t(1-\phi_{Z_1}(-1))}=1.
\end{equation}
Indeed, by l'Hopital's rule we have
\begin{align}
\lim_{t\to\infty}\frac{1}{t}\int_0^t\Big(1-\phi_{Z_1}(-H(u))\Big)\,\mathrm{d}u&=1-\lim_{t\to\infty}\frac{1}{t}\int_0^t\phi_{Z_1}(-H(u))\,\mathrm{d}u\nonumber\\
&=1-\phi_{Z_1}(-1).\nonumber
\end{align}
So \eqref{eq:limg} follows if we check that the second term in \eqref{eq:secondbounded}
is bounded. By the elementary inequality $\mathrm{e}^x\geq 1+x$, $x\in\R$, we have
\begin{align}
0&\leq\int_t^\infty\Big(1-\phi_{Z_1}(H(u-t)-H(u))\Big)\,\mathrm{d}u\nonumber\\
&\leq\EE[Z_1]\int_t^\infty(H(u)-H(u-t))\,\mathrm{d}u\nonumber\\
&\leq\EE[Z_1]\int_0^\infty\left(1-H(u)\right)\,\mathrm{d}u=\EE[Z_1]\int_0^\infty u h(u)\,\mathrm{d}u<\infty,\nonumber
\end{align}
where the latter equality is a consequence of the fact that $h$ is a probability density on $(0,\infty)$.
Finally, we check that \eqref{eq:limg} implies the second limit in \eqref{eq:HYP1} (the first limit being obvious by the definition of $f$).
Letting $o(1)$ denote a function which tends to zero as $y\to\infty$, by \eqref{eq:fx} we have
\begin{align}
\lim_{y\to\infty}\frac{g(y)-g(y+xf(y))}{\sqrt{g(y+xf(y))}}&=\lim_{y\to\infty}\frac{y(1-\phi_{Z_1}(-1))-(y+xf(y))(1-\phi_{Z_1}(-1))+o(1)}{\sqrt{(y+xf(y))(1-\phi_{Z_1}(-1))+o(1)}}
\nonumber\\
&=-\lim_{y\to\infty}\frac{xf(y)(1-\phi_{Z_1}(-1))+o(1)}{\sqrt{(y+xf(y))(1-\phi_{Z_1}(-1))+o(1)}}\nonumber\\
&=-\lim_{y\to\infty}\frac{x(1-\phi_{Z_1}(-1))\sqrt{y}+o(1)}{\sqrt{\lambda(1-\phi_{Z_1}(-1))}\sqrt{(y+xf(y))(1-\phi_{Z_1}(-1))+o(1)}}\nonumber\\
&=-\lim_{y\to\infty}\frac{x(1-\phi_{Z_1}(-1))\sqrt{y}}{\sqrt{\lambda (1-\phi_{Z_1}(-1))}\sqrt{(1-\phi_{Z_1}(-1))y}}=-\frac{x}{\sqrt\lambda}.\nonumber
\end{align}
\noindent$\square$

The proof of Proposition \ref{prop:normalcaching} uses Lemma \ref{prop:normal} below,
which is of its own interest.
We denote by $\mathrm{Lip}(1)$ the class of real-valued Lipschitz functions
from $\R$ to $\R$ with Lipschitz constant less than or equal to
one. Given two real-valued random variables $U$ and $U'$, the
Wasserstein distance between the laws of $U$ and $U'$, written
$d_W(U,U')$, is defined as
\[
d_W(U,U'):=\sup_{\varphi\in\mathrm{Lip}(1)}|\mathbb{E}[\varphi(U)]-\mathbb{E}[\varphi(U')]|.
\]
We recall that the topology induced by $d_W$ on the class of
probability measures over $\R$ is finer than the topology of weak
convergence (see e.g. \cite{nourdin}).
\begin{Lemma}\label{prop:normal}
If $g(t)<\infty$, then
\begin{equation*}
d_W\left(\frac{X(t)-\lambda
g(t)}{\sqrt{\lambda
g(t)}},\mathcal{N}(0,1)\right)\leq\frac{1}{\sqrt{\lambda g(t)}}.
\end{equation*}
\end{Lemma}

\noindent{\bf Remark\,\,3.}
Lemma \ref{prop:normal} provides a Gaussian approximation for $X(t)$ in the Wasserstein distance. On the other hand, Proposition 1 in~\cite{Roberts12} provides
a Gaussian approximation, in the Kolmogorov distance $d_{Kol}$, for the corresponding quantity under the IRM. We note that a Gaussian approximation of $X(t)$, in the Kolmogorov distance, under the SNM
may be easily obtained by Lemma \ref{prop:normal} using the relation
\[
d_{Kol}(X,\mathcal{N}(0,1)):=\sup_{x\in\R}|\mathbb P(X\leq x)-\mathbb P(\mathcal{N}(0,1)\leq x)|\leq 2\sqrt{d_W(X,\mathcal{N}(0,1))},
\]
where $X$ is a real-valued random variable, see e.g. \cite{nourdin}.
\\
\noindent$\square$

\noindent${\it Proof\,\,of\,\,Proposition\,\,\ref{prop:normalcaching}}$.
By the assumptions on $g$ we have $C=\lambda g(t_C)$, $t_C\uparrow\infty$ and $g(t_C)\uparrow\infty$, as $C\uparrow\infty$. For any $x\in\mathbb R$,
\begin{align}
\PP(T_C-t_C>xf(t_C))&=\PP(X(t_C+xf(t_C))<C)\nonumber\\
&=\PP\left(\frac{X(t_C+xf(t_C))-\lambda g(t_C+xf(t_C))}{\sqrt{\lambda g(t_C+xf(t_C))}}<\sqrt{\lambda}\frac{g(t_C)-g(t_C+xf(t_C))}{\sqrt{g(t_C+xf(t_C))}}\right).
\label{eq:normT}
\end{align}
By Lemma \ref{prop:normal} we have
\[
\frac{X(t)-\lambda
g(t)}{\sqrt{\lambda g(t)}}
\overset{law}\longrightarrow
\mathcal{N}(0,1),\quad\text{as $t\to\infty$.}
\]
So, letting $C$ tend to infinity in \eqref{eq:normT} and using \eqref{eq:HYP1} we deduce
\[
\lim_{C\to\infty}\PP\left(\frac{T_C-t_C}{f(t_C)}>x\right)=\PP(\mathcal{N}(0,1)\leq-x)=\PP(\mathcal{N}(0,1)>x).
\]
\noindent$\square$

\noindent${\it Proof\,\,of\,\,Lemma\,\,\ref{prop:normal}}$. Define the Borel measure
$\mu(\mathrm{d}x):=\lambda\mathrm{d}g(x)$ over $[0,t]$ (note that
$g$ increases on $[0,t]$ and so $\mathrm{d}g$ is a
Lebesgue-Stieltjes measure) and the function
$h(x):=\ind_{[0,t]}(x)/\sqrt{\lambda g(t)}$, $x\in [0,t]$. By
Corollary 3.4 in \cite{peccati} and Proposition \ref{prop:lap}, we have
\begin{align}
d_W\left(\frac{X(t)-\lambda g(t)}{\sqrt{\lambda g(t)}},\mathcal{N}(0,1)\right)\leq
\Big|1-\int_{[0,t]}|h(x)|^2\,\mu(\mathrm{d}x)\Big|+\int_{[0,t]}|h(x)|^3\,\mu(\mathrm{d}x)
=\frac{1}{\sqrt{\lambda g(t)}}.\nonumber
\end{align}
\noindent$\square$

\subsection{The \lq\lq in" and the ``hit" probabilities}
The results of the previous subsection provide a justification to the Che approximation, as $C\to\infty$. Indeed, the law of large numbers
\eqref{eq:cheas} guarantees that, asymptotically in $C$, $t_C$ is a correct approximation of $T_C$; the bounds
\eqref{eq:LD1rev} and \eqref{eq:LD2rev} guarantee that deviations of $T_C$ from its most probable value $t_C$ are, asymptotically in $C$, exponentially small;
the normal approximation \eqref{varsmall} allows us to identify the typical asymptotic values of $T_C$ via the construction of asymptotic confidence intervals.
In this subsection we provide complementary non-asymptotic analytical upper bounds on the prediction error
entailed by Che's approximation of the \lq\lq hit'' probability (see Proposition \ref{prop:error0}).
This result allows us to assess the accuracy of the Che approximation in many cases of practical interest, cf. Section \ref{sec:numerical}.

\subsubsection{The \lq\lq in" probability}

The "in" probability is defined as the probability of finding at time $t$ a tagged content $m_0$ in the cache, given that
$\xi_{m_0}=x$ and $Z_{m_0}=z$. Thus:\\
\noindent $(i)$ Under Che's approximation, the "in" probability is given by
\begin{align}
p_{\mathrm{in},\mathrm{Che}}^{(t-x)}(z,t_C):&=\PP(N^{(m_0)}((t-t_C,t])\geq 1\,\mid\,(\xi_{m_0},Z_{m_0})=(x,z))\nonumber\\
&=1-\mathrm{e}^{-\int_{t-x-t_C}^{t-x}h(u,z)\,\mathrm{d}u},\label{eq:pinche}
\end{align}
where $N^{(m_0)}(A)$ denotes the number of points
$\{T_n^{(m_0)}\}_{n\geq 1}$ in $A\subset\mathbb R$;\\
\noindent $(ii)$
Without relying on Che's approximation, the conditional "in" probability is given by
\begin{align}
p_{\mathrm{in}}^{(t-x)}(z,T_C):&=\PP(N^{(m_0)}((t-T_C,t])\geq 1\,|\,(\xi_{m_0},Z_{m_0})=(x,z),T_{C})\nonumber\\
&=p_{\mathrm{in},\mathrm{Che}}^{(t-x)}(z,T_C).\nonumber
\end{align}

\subsubsection{The ``hit" probability}

The ``hit" probability is defined as the ratio between the average
rate at which "hits" of a tagged content occur and the average rate at which requests of the tagged content are observed. Thus:\\
\noindent $(iii)$ Under Che's approximation, the ``hit" probability is given by
\begin{align}
p_{\mathrm{hit},\mathrm{Che}}(t_C):&=\frac{\EE[h(t-\xi_{m_0},Z_{m_0})p^{(t-\xi_{m_0})}_{\mathrm{hit},\mathrm{Che}}(Z_{m_0},t_C)]}
{\EE[h(t-\xi_{m_0},Z_{m_0})]}\nonumber\\
&=\frac{\EE[h(t-\xi_{m_0},Z_{m_0})p^{(t-\xi_{m_0})}_{\mathrm{in},\mathrm{Che}}(Z_{m_0},t_C)]}
{\EE[h(t-\xi_{m_0},Z_{m_0})]},\label{eq:refdonoyund}
\end{align}
with the convention $0/0=0$. The equality \eqref{eq:refdonoyund} is a consequence of the fact that,
under Che's approximation, the probability (denoted by $p_{\mathrm{hit},\mathrm{Che}}^{(t-x)}(z,t_C)$) that the tagged content $m_0$,
introduced into the catalogue at time $\xi_{m_0}=x$ and with mark $Z_{m_0}=z$, is found in the cache by an arriving request at time $t$ is equal to $p_{\mathrm{in},\mathrm{Che}}^{(t-x)}(z,t_C)$. Indeed
\begin{align}
p_{\mathrm{hit},\mathrm{Che}}^{(t-x)}(z,t_C):&=\PP\Big(\sum_{T_n^{(m_0)}\in N^{(m_0)}\setminus\{t\}}\ind_{(t-t_C,t]}(T_n^{(m_0)})\ge 1\,\Big|\,t\in N^{(m_0)},\,
(\xi_{m_0},Z_{m_0})=(x,z)\Big)\nonumber\\
&=\PP\Big(N^{(m_0)}((t-t_C,t])\geq 1)\Big| (\xi_{m_0},Z_{m_0})=(x,z)\Big)\label{eq:Sliv}\\
&=p^{(t-x)}_{\mathrm{in},\mathrm{Che}}(z,t_C),\nonumber
\end{align}
where the equality \eqref{eq:Sliv} is a consequence of the Slivnyak-Mecke theorem (see e.g. Proposition 13.1.VII p. 281 in \cite{daley2}).

Note that the probability $p_{\mathrm{hit},\mathrm{Che}}(t_C)$ does not depend on $m_0$ and $t$. Indeed, for an arbitrary $s<t$ we have
\begin{align}
\frac{\EE[h(t-\xi_{m_0},Z_{m_0})p_{\mathrm{in},\mathrm{Che}}^{(t-\xi_{m_0})}(Z_{m_0},t_C)\ind\{s<\xi_{m_0}<t\}]}{\EE[h(t-\xi_{m_0},Z_{m_0})\ind\{s<\xi_{m_0}<t\}]}
&=\frac{(t-s)^{-1}\int_{s}^{t}\EE[h(t-u,Z_{1})p_{\mathrm{in},\mathrm{Che}}^{(t-u)}(Z_{1},t_C)]\,\mathrm{d}u}{(t-s)^{-1}\int_{s}^{t}\EE[h(t-u,Z_{1})]\,\mathrm{d}u}\nonumber\\
&=\frac{\int_{s}^{t}\EE[h(t-u,Z_{1})p_{\mathrm{in},\mathrm{Che}}^{(t-u)}(Z_{1},t_C)]\,\mathrm{d}u}{\int_{s}^{t}\EE[h(t-u,Z_{1})]\,\mathrm{d}u},\nonumber
\end{align}
and so letting $s$ tend to $-\infty$ we deduce
\begin{align}
p_{\mathrm{hit},\mathrm{Che}}(t_C)=
\frac{\int_{0}^{\infty}\EE[h(u,Z_{1})p_{\mathrm{in},\mathrm{Che}}^{(u)}(Z_{1},t_C)]\,\mathrm{d}u}{\int_{0}^{\infty}\EE[h(u,Z_{1})]\,\mathrm{d}u};\label{phitche}
\end{align}
\noindent $(iv)$ Without relying on Che's approximation, the conditional ``hit" probability is given by
\begin{align}
p_{\mathrm{hit}}(T_C):=\frac{\EE[h(t-\xi_{m_0},Z_{m_0})p^{(t-\xi_{m_0})}_{\mathrm{in},\mathrm{Che}}(Z_{m_0},T_C)\,|\,T_C]}
{\EE[h(t-\xi_{m_0},Z_{m_0})]}.\nonumber
\end{align}
Arguing as above, one may easily check that
\begin{align}
p_{\mathrm{hit}}(T_C)&=\frac{\int_{0}^{\infty}\EE[h(u,Z_{m_0})p_{\mathrm{in},\mathrm{Che}}^{(u)}(Z_{m_0},T_C)\mid T_C]\,\mathrm{d}u}{\int_{0}^{\infty}\EE[h(u,Z_{1})]\,\mathrm{d}u}.\nonumber
\end{align}
Being $Z_{m_0}$ and $T_C$ independent, the (unconditional) ``hit" probability is given by
\begin{align}
p_{\mathrm{hit}}&
=\int_{[0,\infty)}p_{\mathrm{hit},\mathrm{Che}}(\theta)\,\PP_{T_C}(\mathrm{d}\theta),\nonumber
\end{align}
where $\PP_{T_C}$ denotes the law of $T_C$ and $p_{\mathrm{hit},\mathrm{Che}}(\theta)$ is defined as $p_{\mathrm{hit},\mathrm{Che}}(t_C)$ with $\theta$ in place of $t_C$.

\subsubsection{Error estimate}

By using the above relations and classical estimates for the tail of a Poisson distribution, we can evaluate the error
committed by approximating $p_{\mathrm{hit}}$ with $p_{\mathrm{hit},\mathrm{Che}}(t_C)$. The following proposition holds.
\begin{Proposition}\label{prop:error0}
If $g:(0,\infty)\to (0,\infty)$ is strictly increasing, then, for any $\delta\in (0,1)$ and $C>0$, we have
\begin{align}
|p_{\mathrm{hit}}-p_{\mathrm{hit},\mathrm{Che}}(t_C)|&\leq\exp(-\lambda g(t_C(1-\delta))R(C/\lambda g(t_C(1-\delta))))\nonumber\\
&\,\,\,\,\,\,\,
\,\,\,\,\,\,\,
\,\,\,\,\,\,\,
+\exp(-\lambda g(t_C(1+\delta))R(C/\lambda g(t_C(1+\delta))))\nonumber\\
&\,\,\,\,\,\,\,
\,\,\,\,\,\,\,
\,\,\,\,\,\,\,
+\max_{\theta\in \{t_C(1-\delta),t_C(1+\delta)\}}|p_{\mathrm{hit},\mathrm{Che}}(\theta)-p_{\mathrm{hit},\mathrm{Che}}(t_C)|,\nonumber
\end{align}
where $R(x):=1-x+x\log x$, $x>0$.
\end{Proposition}

This proposition allows an assessment of the accuracy of Che's approximation in different scenarios.
As shown by the numerical simulations in \cite{nostro-tec-rep}, see also Section~\ref{sec:numerical}, in most cases by
exploiting Proposition \ref{prop:error0} we can show that Che's approximation
leads to surprisingly accurate predictions of caching performance.

\noindent{\it Proof\,\,of\,\,Proposition\,\,\ref{prop:error0}.} We preliminary note that, for any $\delta\in (0,1)$ and $C>0$, we have
\begin{equation}\label{eq:ineqpenrose}
\lambda g(t_C(1-\delta))\leq C\leq\lambda g(t_C(1+\delta)).
\end{equation}
Indeed, since $g$ is strictly increasing \eqref{eq:ineqpenrose} is equivalent to
$t_C(1-\delta)\leq g^{-1}(C/\lambda)\leq t_C(1+\delta)$, which holds since $t_C=g^{-1}(C/\lambda)$.
Note that, due to \eqref{eq:pinche}, $p_{\mathrm{hit},\mathrm{Che}}(\cdot)$ is a non-decreasing function.
So, for all $\delta\in (0,1)$, we have
\begin{align}
|p_{\mathrm{hit}}-p_{\mathrm{hit},\mathrm{Che}}(t_C)|
&\leq\int_{[0,t_C(1-\delta)]}(p_{\mathrm{hit},\mathrm{Che}}(t_C)-p_{\mathrm{hit},\mathrm{Che}}(\theta))\,\PP_{T_C}(\mathrm{d}\theta)\nonumber\\
&
\,\,\,\,\,\,
\,\,\,\,\,\,
+\int_{(t_C(1+\delta),\infty)}(p_{\mathrm{hit},\mathrm{Che}}(\theta)-p_{\mathrm{hit},\mathrm{Che}}(t_C))\,\PP_{T_C}(\mathrm{d}\theta)\nonumber\\
&
\,\,\,\,\,\,
\,\,\,\,\,\,
+\int_{(t_C(1-\delta),t_C(1+\delta)]}|p_{\mathrm{hit},\mathrm{Che}}(\theta)-p_{\mathrm{hit},\mathrm{Che}}(t_C)|\,\PP_{T_C}(\mathrm{d}\theta)\nonumber\\
&\leq\PP(T_C\leq t_C(1-\delta))+\PP(T_C> t_C(1+\delta))\nonumber\\
&\,\,\,\,\,\,
\,\,\,\,\,\,
+\max_{\theta\in\{t_C(1-\delta),t_C(1+\delta)\}}|p_{\mathrm{hit},\mathrm{Che}}(\theta)-p_{\mathrm{hit},\mathrm{Che}}(t_C)|.\nonumber
\end{align}
The claim follows noticing that by the definition of $T_C$, the inequality \eqref{eq:ineqpenrose} and
the properties of the Poisson distribution (see e.g. Lemma 1.2 in \cite{penrose}, formulas
(1.10) and (1.11)) we have
\begin{align}
\PP(T_C\leq t_C(1-\delta))=\PP(X(t_C(1-\delta))>C)\leq\exp(-\lambda g(t_C(1-\delta))R(C/\lambda g(t_C(1-\delta))))\nonumber
\end{align}
and
\begin{align}
\PP(T_C>t_C(1+\delta))=\PP(X(t_C(1+\delta))\leq C)\leq\exp(-\lambda g(t_C(1+\delta))R(C/\lambda g(t_C(1+\delta)))).\nonumber
\end{align}
\noindent$\square$

\subsection{Extension to the case of contents with variable sizes}
At a first glance, dealing with contents of variable sizes may appear significantly more challenging. Indeed, before inserting in the cache a new content,
enough memory must be freed by selecting a proper set of objects to expunge.
The  content to be stored, then,  need to be  partitioned into  small
portions (fragments) that fit into the non-adjacent areas of memory, each one corresponding to a different fragment of the  expunged contents. Unfortunately an excessive
fragmentation of the contents can significantly reduce the bandwidth performance (speed) of the
cache and therefore must be prevented by executing complex memory management operations such as periodic de-fragmentation.
A simple method to
cache contents of variable sizes, referred in the following as {\it chunkization}, consists in breaking each content into an integer number
of pieces with a fixed size, called {\it chunks}, which are treated as independent objects by the caching system.
By properly dimensioning the size of the  chunk it is  possible to achieve an optimal trade-off between memory efficiency and bandwidth performance.
Indeed, by enlarging the size of the chunk, memory efficiency decreases (for the effect of the last chunk size rounding), while the
cache speed increases since the size of fragments (which are memorized in consecutive memory locations) increases.
In this way the degradation of the cache due to content fragmentation is kept under control, without the necessity of executing complex memory management operations.
This is the main reason why chunkization has become an almost universally adopted technique in caching systems supporting the distribution of contents over the Internet
\cite{Jacobson-ICN, ICN-1,chunking}.

In this subsection we briefly discuss how our approach can be extended to evaluate the effectiveness of Che's approximation for an
LRU cache which stores contents of variable sizes through chunkization.

We still assume that the LRU cache operates under the SNM: requests of different chunks corresponding to the same content $m$ are perfectly synchronized,
and the process of requests for each chunk of content $m$ is a Cox process with stochastic intensity $\lambda_m(t)$. We denote by $A_m$ the number of chunks
in which the content $m$
is partitioned and assume that $\{A_m\}_{m\geq 1}$ is a sequence of independent and identically distributed random variables with values on $\{1,2,\ldots\}$,
independent of $\{\xi_m\}_{m\geq 1}$ and $\{Z_m\}_{m\geq 1}$.
The number of  chunks (corresponding to contents different from $m_0$) requested in the time interval
$[0,t]$ is given by
\[
\bold{X}_{m_0}(t):=\sum_{m\neq m_0}A_m\ind\{m \mbox{ requested in } [0,t],\,\xi_m\in (-\infty,t]\}.
\]
Setting $X_{m_0,x}(t):=(\bold{X}_{m_0}(t)\,|\,\xi_{m_0}=-x)+A_{m_0}-1$, with $x>0$, we define the cache eviction time as
\[
T_C(m_0,x):=\inf\{t\geq 0:\,\,X_{m_0,x}(t)=C\}=\inf\{t\geq 0:\,\, (\bold{X}_{m_0}(t)\,|\,\xi_{m_0}=-x) =C-A_{m_0}+1\},
\]
where we express the caching storage capacity $C$ in number of  chunks.
The definition of $T_C(m_0,x)$ reflects the fact that we consider a content to be expunged
(i.e. unavailable at the cache) when its first chunk is expunged by the cache.  

Let $g$ be the function defined by \eqref{eq:grev}.
If $g(t)<\infty$ and the $A$'s are light-tail, i.e.
\begin{equation}\label{eq;lightail}
\text{$\exists$ a right neighborhood of zero, say $\mathcal{N}_+$, such that
$\phi_{A_1}(\theta):=\mathbb{E}[\mathrm{e}^{\theta A_1}]<\infty$ $\forall$ $\theta\in\mathcal{N}_+$,}
\end{equation}
then, arguing as in the proof of Proposition \ref{prop:lap}, one has that $\bold{X}_{m_0}(t)\,|\,\xi_{m_0}=-x$ follows the same law of $\sum_{i=1}^{S}A_i$,
where $S$ is a Poisson distributed random variable with mean $\lambda g(t)$ and $S$ is independent of $\{A_m\}_{m\geq 1}$.
Note that the laws of $X_{m_0,x}(t)$ and $T_C(m_0,x)$ do not depend on $x$, but they depend on $m_0$. However, for ease of notation, hereafter
we omit to explicitly indicate this dependence, writing $X(t)$ and $T_C$ in place of $X_{m_0,x}(t)$ and $T_C(m_0,x)$. In this context, Che's approximation
of $T_C$ is
\[
t_C:=\inf\{t\geq 0:\,\,\mathbb{E}[X_{m_0,x}(t)]=C\}=
\inf\{t\geq 0:\,\,\lambda g(t)=(C+1-\mathbb{E}[A_1])/\mathbb{E}[A_1]\}.
\]
Under the same assumptions of Proposition
\ref{cor:ldp} and condition \eqref{eq;lightail}, we have that the family of random variables $\{g(T_C)/C\}_{C\geq 1}$
obeys a large deviation principle on $[0,\infty)$ with speed $v(C):=C$ and rate function $I(x):=x\Lambda^*(1/x)$, $x>0$, $I(0):=\lim_{x\to 0^+}x\Lambda^*(1/x)$, where
\[
\Lambda^*(x):=\sup_{\theta\in\mathbb R}(\theta x-\lambda(\mathbb{E}[\mathrm{e}^{\theta A_1}]-1)).
\]
Since the derivation of this large deviation principle is not immediate, we sketch the proof.
Arguing as in the proof of Lemma \ref{prop:ldp} one has that the process $\{\tilde{X}(t)/g(t)\}_{t\geq 1}$, where $\tilde{X}(t):=\bold{X}_{m_0}(t)\,|\,\xi_{m_0}=-x$, obeys a large deviation principle
on $[0,\infty)$ with speed $g$ and rate function $I_1(u):=\Lambda^*(u)$. On the other hand, by using the definition of large deviation principle it is readily checked
that the process $\{(A_{m_0}-1)/g(t)\}_{t\geq 1}$ obeys a large deviation principle on $[0,\infty)$
with speed $g$ and rate function $I_2(v):=+\infty\ind\{v>0\}$, with the convention $\infty\cdot 0=0$. By the independence of the processes
$\{\tilde{X}(t)/g(t)\}_{t\geq 1}$ and $\{(A_{m_0}-1)/g(t)\}_{t\geq 1}$ and the contraction principle (i.e. by Exercise 4.2.7 on p. 129 and Theorem 4.2.1 on p. 126 in \cite{dembo})
one has that the process $\{X(t)/g(t)\}_{t\geq 1}$ obeys a large deviation principle
on $[0,\infty)$ with speed $g$ and rate function
\[
\inf\{I_1(u)+I_2(v):\,\,u+v=x\}=I_1(x)=\Lambda^*(x).
\]
The claimed large deviation principle for the family of random variables $\{g(T_C)/C\}_{C\geq 1}$ follows by applying Theorem \ref{thm:duffield}
as in the proof of Lemma \ref{prop:ldp} (note that $\{T_C\}$ is the inverse process of $\{X(t)\}$).

By this large deviation principle one can obtain the law of large numbers \eqref{eq:cheas},
the tail estimates \eqref{eq:tail}, \eqref{eq:tail1} and the deviation bounds \eqref{eq:LD1rev}, \eqref{eq:LD2rev},
cf. the proofs of Propositions \ref{cor:lln}, \ref{cor:ldp} and Corollary \ref{cor:rem1}, respectively.

Finally, we note that the proofs of Propositions \ref{prop:normalcaching} and \ref{prop:error0} may be easily adapted
in order to obtain a normal approximation of the cache eviction time and an estimate of the
error committed by approximating the corresponding \lq\lq hit\rq\rq probability with its expression under Che's approximation, we omit the details.

\section{Networks of caches: the case of two caches is series}

The analysis of networks of caches 
is a difficult task, indeed
an exact characterization of the miss stream of an LRU cache is in general prohibitive.
Under the IRM a standard and rather crude approach proposed in the literature (see e.g. \cite{kurose}) 
consists in: $i)$ approximating the miss stream of a content at a cache with a homogeneous Poisson process
whose rate matches the miss stream rate; $ii)$ assuming the state of caches to be independent.
However, significant errors may be experienced. An alternative approach, that has been recently proposed
for feed-forward networks of LRU caches (such as networks with linear topologies or trees)
consists in approximating  the real miss stream  with that of a cache operating under Che's approximation,
(see \cite{bianchi} and \cite{towsley-nain}). This approach has been experimentally shown to be potentially fairly accurate, but, unfortunately,
at the same time, it is computationally highly expensive~\cite{towsley-nain}. Recently a more efficient procedure has been proposed in~\cite{bianchi},
where further approximations are considered to simplify the computation of the 	\lq\lq hit'' probability. However, in this latter case, the accuracy of the estimate is in part sacrificed.

Here we show how the approach of~\cite{towsley-nain} can be adapted to the SNM.
Our study reveals that the exact computation of the ``hit'' probability, under Che's  approximation,
for a simple tandem network of caches (i.e. a network of two LRU  caches in series)
is computationally hard (see Remark 4). This is mainly due to the effect
of the complex dependencies between the states of the two caches.

Since the analysis at the first cache can be carried on as in the previous section, here we focus on the second cache. Note
that an arriving request for content $m_0$ can produce a ``hit''  at the second cache only if
it misses the content $m_0$ at the first cache. 
So, under Che's approximation,
the ``hit'' probability for content  $m_0$,
introduced into the catalogue at time $\xi_{m_0}=x$ and with mark $Z_{m_0}=z$,
 is given by
\begin{align}
&p_{\mathrm{hit},\mathrm{Che},\mathrm{II}}^{(t-x)}(z,t_{C_1},t_{C_2}):=\PP\Big(\sum_{T_n^{(m_0)}\in N^{(m_0)}\setminus\{t\}}\ind_{(t-t_{C_1},t]}(T_n^{(m_0)})=0,\nonumber\\
&\,\,\,\,\,\,\,\,\,\,\,\,\,
\sum_{n}\ind_{(t-t_{C_2},t)}(T_n^{(m_0)})\ind\{T_n^{(m_0)}-T_{n-1}^{(m_0)}>t_{C_1}\}\ge 1\,\Big|\,t\in N^{(m_0)},
(\xi_{m_0},Z_{m_0})=(x,z)\Big),\label{eq:phittandem}
\end{align}
where $t_{C_i}$ denotes the cache eviction time at the cache $i\in\{1,2\}$ under Che's approximation.

Hereafter, the symbol $\sum_{i_1<i_2}^{0,k-1}$ denotes the sum over all the couples $(i_1,i_2)\in\{0,\ldots,k-1\}^2$ such that $i_1<i_2$.
The following proposition holds.
\begin{Proposition}\label{prop:tandem}
We have that
\begin{equation}\label{pzero}
p_{\mathrm{hit},\mathrm{Che},\mathrm{II}}^{(t-x)}(z,t_{C_1},t_{C_2})=0\quad\text{if $t_{C_2}\leq t_{C_1}$}
\end{equation}
and
\begin{equation}\label{UL}
\mathfrak{L}\leq p_{\mathrm{hit},\mathrm{Che},\mathrm{II}}^{(t-x)}(z,t_{C_1},t_{C_2})\leq\mathfrak{U}
\quad\text{if, for some integer $k\geq 1$, $kt_{C_1}<t_{C_2}\leq (k+1)t_{C_1}$.}
\end{equation}
Here
\begin{align}
\mathfrak{L}:=&\mathrm{e}^{-\int_{t-x-t_{C_{1}}}^{t}h(s,z)\,\mathrm{d}s}\Big(
\int_{t-x-t_{C_2}}^{t-x-t_{C_1}}h(\tau,z)\mathrm{e}^{-\int_{\tau-t_{C_{1}}}^{\tau}h(s,z)\,\mathrm{d}s}\mathrm{d}\tau
\nonumber\\
&\,\,\,\,\,\,\,
\,\,\,\,\,\,\,
-\sum_{i_1<i_2}^{0,k-1}
\int_{b_{i_1}-x}^{b_{i_1+1}-x}h(\tau,z)\mathrm{e}^{-\int_{\tau-t_{C_{1}}}^{\tau}h(s,z)\,\mathrm{d}s}\mathrm{d}\tau
\int_{b_{i_2}-x}^{b_{i_2+1}-x}h(\tau,z)\mathrm{e}^{-\int_{\tau-t_{C_{1}}}^{\tau}h(s,z)\,\mathrm{d}s}\mathrm{d}\tau\Big),
\nonumber
\end{align}
\[
\mathfrak U:=\mathrm{e}^{-\int_{t-x-t_{C_{1}}}^{t}h(s,z)\,\mathrm{d}s}
\int_{t-x-t_{C_2}}^{t-x-t_{C_1}}h(\tau,z)\mathrm{e}^{-\int_{\tau-t_{C_{1}}}^{\tau}h(s,z)\,\mathrm{d}s}\mathrm{d}\tau\nonumber
\]
and
\[
b_i:=\frac{(k-i)(t-t_{C_2})+i(t-t_{C_1})}{k},\quad i=0,\ldots,k.
\]
\end{Proposition}

\noindent{\it Proof.}
By \eqref{eq:phittandem} and the Slivnyak-Mecke theorem (see e.g. Proposition 13.1.VII p. 281 in \cite{daley2}), we have
\begin{align}
&p_{\mathrm{hit},\mathrm{Che},\mathrm{II}}^{(t-x)}(z,t_{C_1},t_{C_2})\nonumber\\
&=\PP\Big(N^{(m_0)}((t-t_{C_1},t])=0,\nonumber\\
&\,\,\,\,\,\,\,\,\,\,\,\,\,
\sum_{n}\ind_{(t-t_{C_2},t-t_{C_1}]}(T_n^{(m_0)})\ind\{T_n^{(m_0)}-T_{n-1}^{(m_0)}>t_{C_1}\}\ge 1\,\Big|\,
(\xi_{m_0},Z_{m_0})=(x,z)\Big),\nonumber
\end{align}
and this quantity is equal to zero if $t_{C_1}\geq t_{C_2}$, which proves \eqref{pzero}.
Otherwise, there exists an integer $k\geq 1$ such that
$kt_{C_1}<t_{C_2}\leq (k+1)t_{C_1}$. We consider the partition of the set $(t-t_{C_2},t-t_{C_1}]$
formed by the intervals $I_i:=(b_i,b_{i+1}]$, $0\leq i\leq k-1$,
where the $b_i$'s are defined in the statement,
and we set $n_i^*:=\min\{n:\,\,T_n^{(m_0)}>b_i\}$. Since, by construction $b_{i+1}-b_i \le t_{C_1}$, for any $ 0\leq i\leq k-1$,   provided that
$T_n^{(m_0)}\in I_{\bar i}$, for some $\bar i=0,\ldots,k$, and $T_n^{(m_0)}-T_{n-1}^{(m_0)}>t_{C_1}$, then necessarily $T_{n-1}^{(m_0)}\leq b_{\bar i}$.
Therefore, setting $A:=\{N^{(m_0)}((t-t_{C_1},t])=0\}$ and
\begin{equation}\label{eq:Bi}
B_i:=\{T_{n_{i}^*}^{(m_0)}\in I_i,\,N^{(m_0)}((T_{n_i^*}^{(m_0)}-t_{C_1},b_i))=0\},\quad i=0,\ldots,k-1
\end{equation}
we deduce
\begin{align}
&p_{\mathrm{hit},\mathrm{Che},\mathrm{II}}^{(t-x)}(z,t_{C_1},t_{C_2})
=\PP\Big(A\cap\Big(\bigcup_{i=0}^{k-1}B_i\Big)\,\Big|\,(\xi_{m_0},Z_{m_0})=(x,z)\Big)\nonumber\\
&\,\,=\PP\Big(A\,\Big|\,(\xi_{m_0},Z_{m_0})=(x,z)\Big)\PP\Big(\bigcup_{i=0}^{k-1}B_i\,\Big|\,(\xi_{m_0},Z_{m_0})=(x,z)\Big)\nonumber\\
&\,\,=\mathrm{e}^{-\int_{t-x-t_{C_{1}}}^{t}h(s,z)\,\mathrm{d}s}\PP\Big(\bigcup_{i=0}^{k-1}B_i\,\Big|\,(\xi_{m_0},Z_{m_0})=(x,z)\Big).\label{eq:hitbound}
\end{align}
The Bonferroni inequality and the union bound yield
\begin{align}
\sum_{i=0}^{k-1}\PP\Big(B_i\,\Big|\,(\xi_{m_0},Z_{m_0})=(x,z)\Big)&-\sum_{i_1<i_2}^{0,k-1}\PP\Big(B_{i_1}\cap B_{i_2}\,|\,(\xi_{m_0},Z_{m_0})=(x,z)\Big)\nonumber\\
&\leq\PP\Big(\bigcup_{i=0}^{k-1}B_i\,\Big|\,(\xi_{m_0},Z_{m_0})=(x,z)\Big)\nonumber\\
&\leq\sum_{i=0}^{k-1}\PP\Big(B_i\,\Big|\,(\xi_{m_0},Z_{m_0})=(x,z)\Big).\label{eq:bounds}
\end{align}
For any $i_1<i_2$, $i_1,i_2\in\{0,\ldots,k-1\}$, we have
\begin{align}
&\PP\Big(B_{i_1}\cap B_{i_2}\,\Big|\,(\xi_{m_0},Z_{m_0})=(x,z)\Big)\nonumber\\
&=\Big(\PP\Big(B_{i_2},T_{n_{i_2}^*}-T_{n_{i_1}^*}>t_{C_1}\,\Big|\,B_{i_1},\,(\xi_{m_0},Z_{m_0})=(x,z)\Big)\nonumber\\
&\,\,\,\,\,\,\,
\,\,\,\,\,\,\,
+\PP\Big(B_{i_2},T_{n_{i_2}^*}-T_{n_{i_1}^*}\leq t_{C_1}\,\Big|\,B_{i_1},\,(\xi_{m_0},Z_{m_0})=(x,z)\Big)\Big)
\PP\Big(B_{i_1}\,\Big|\,(\xi_{m_0},Z_{m_0})=(x,z)\Big)\nonumber\\
&=\PP\Big(B_{i_2},T_{n_{i_2}^*}-T_{n_{i_1}^*}>t_{C_1}\,\Big|\,B_{i_1},\,(\xi_{m_0},Z_{m_0})=(x,z)\Big)
\PP\Big(B_{i_1}\,\Big|\,(\xi_{m_0},Z_{m_0})=(x,z)\Big),\nonumber
\end{align}
where the latter equality follows noticing that, given $B_{i_1}$, $\{B_{i_2},T_{n_{i_2}^*}-T_{n_{i_1}^*}\leq t_{C_1}\}=\emptyset$. By the independence of the increments
of the Poisson process we deduce
\begin{align}
\PP\Big(B_{i_2},T_{n_{i_2}^*}-T_{n_{i_1}^*}>t_{C_1}\,\Big|\,B_{i_1},\,(\xi_{m_0},Z_{m_0})=(x,z)\Big)
=\PP\Big(B_{i_2},T_{n_{i_2}^*}-T_{n_{i_1}^*}>t_{C_1}\,\Big|\,(\xi_{m_0},Z_{m_0})=(x,z)\Big),\nonumber
\end{align}
and so
\begin{align}
\PP\Big(B_{i_1}\cap B_{i_2}\,\Big|\,(\xi_{m_0},Z_{m_0})=(x,z)\Big)
\leq\PP\Big(B_{i_1}\,\Big|\,(\xi_{m_0},Z_{m_0})=(x,z)\Big)\PP\Big(B_{i_2}\,\Big|\,(\xi_{m_0},Z_{m_0})=(x,z)\Big).\nonumber
\end{align}
Consequently, by \eqref{eq:hitbound} and \eqref{eq:bounds}
we have the following bounds on the ``hit" probability
\begin{align}
&\mathrm{e}^{-\int_{t-x_{m_0}-t_{C_{1}}}^{t}h(s,z)\,\mathrm{d}s}\Big(\sum_{i=0}^{k-1}\PP\Big(B_i\,\Big|\,(\xi_{m_0},Z_{m_0})=(x,z)\Big)\nonumber\\
&\,\,\,\,\,\,\,
\,\,\,\,\,\,\,
-\sum_{i_1<i_2}^{0,k-1}\PP\Big(B_{i_1}\,|\,(\xi_{m_0},Z_{m_0})=(x,z)\Big)\PP\Big(B_{i_2}\,|\,(\xi_{m_0},Z_{m_0})=(x,z)\Big)\Big)
\nonumber\\
&\leq p_{\mathrm{hit},\mathrm{Che},\mathrm{II}}^{(t-x)}(z,t_{C_1},t_{C_2})\nonumber\\
&\leq\mathrm{e}^{-\int_{t-x-t_{C_{1}}}^{t}h(s,z)\,\mathrm{d}s}
\sum_{i=0}^{k-1}\PP\Big(B_i\,\Big|\,(\xi_{m_0},Z_{m_0})=(x,z)\Big).\label{prelbd}
\end{align}
Relation \eqref{UL} follows by \eqref{prelbd} and the following computation:
\begin{align}
&\PP\Big(B_i\,\Big|\,(\xi_{m_0},Z_{m_0})=(x,z)\Big)\nonumber\\
&=
\int_{b_i}^{b_{i+1}}\PP(N^{(m_0)}((\tau-t_{C_1},b_i))=0\,\mid\,(\xi_{m_0},Z_{m_0})=(x,z))\,\PP_{T_{{n_i}^*}^{(m_0)}\,|\,(\xi_{m_0},Z_{m_0})=(x,z)}(\mathrm{d}\tau)
\nonumber\\
&=\int_{b_i}^{b_{i+1}}\mathrm{e}^{-\int_{\tau-t_{C_{1}}}^{b_i}h(s-x,z)\,\mathrm{d}s}\,\PP_{T_{{n_i}^*}^{(m_0)}\,|\,(\xi_{m_0},Z_{m_0})=(x,z)}(\mathrm{d}\tau)
\nonumber\\
&=\int_{b_i}^{b_{i+1}}\mathrm{e}^{-\int_{\tau-t_{C_{1}}}^{b_i}h(s-x,z)\,\mathrm{d}s}\,h(\tau-x,z)\mathrm{e}^{-\int_{b_i}^{\tau}h(s-x,z)\,\mathrm{d}s}\mathrm{d}\tau
\nonumber\\
&=
\int_{b_i-x}^{b_{i+1}-x}h(\tau,z)\mathrm{e}^{-\int_{\tau-t_{C_{1}}}^{\tau}h(s,z)\,\mathrm{d}s}\mathrm{d}\tau.
\nonumber
\end{align}
\noindent$\square$

\noindent{\bf Remark\,\,4.}
Proposition~\ref{prop:tandem} provides the exact value of the ``hit" probability
when $t_{C_2}\leq 2t_{C_1}$. As it is clear from the proof, one might exactly compute the ``hit" probability
even when $t_{C_2}>2t_{C_1}$ by applying the inclusion-exclusion formula. However, the resulting
computational cost would be very high since one has to compute the
probability of any intersection of the events $B_i$ defined by \eqref{eq:Bi}.
In conclusion, we can say that any computationally efficient approach to the performance analysis of a
tandem network of caches must resort to some extra approximations (in addition to Che's approximation), which affect
inevitably the accuracy of the method.
\\
\noindent{\bf Remark\,\,5.}
In principle, the approach proposed in this section can be generalized to networks of caches with a feed-forward structure
i.e., roughly speaking, to networks of caches in which the caches are \lq\lq ordered\rq\rq (in some sense) and any content request follows a path
that traverses   caches  in \lq \lq increasing order\rq\rq .
Typical examples are networks of caches with a tree structure, where
a request for a generic content follows a path in the network which starts from a cache placed on a leaf (belonging conventionally to the level 1 of the tree), it is directed toward 
the cache placed at the root
(belonging conventionally to the level $K$ of the tree) 
and stops as soon as a \lq\lq hit\rq\rq is produced.
Note that an arriving request for the content $m_0$
can produce a \lq\lq hit\rq\rq $\,$ at a cache located at the $k$th level only if it has been missed at caches located at the $i$th level,
for any $i\leq k-1$.
So, applying similar arguments as in \eqref{eq:phittandem}, one may obtain a formal expression for the probability that there is a \lq\lq hit\rq\rq $\,$at a cache located
at the $k$th level of the tree. However, the numerical evaluation of this probability becomes more and more prohibitive as the level grows, i.e. as $k$
increases.
As for the case of tandem networks of caches, any computationally efficient approach
must rely on some additional approximations (in addition to Che's approximation)
which reduce the accuracy of the method. Recently, several approximations have been proposed
\cite{bianchi,towsley-nain,nostro_infocom,kurose,nostro-TMM}. The accuracy of such approximations
varies significantly from scenario to scenario and can be evaluated experimentally
by comparing analytical predictions against Monte Carlo simulations~\cite{bianchi,towsley-nain,nostro_infocom,kurose,nostro-TMM}.


\section{Numerical illustrations}\label{sec:numerical}


As shown by several recent experimental works,
many video contents (such as YouTube contents)
exhibits few typical normalized temporal popularity profiles, each profile corresponding
to a large class of contents with similar characteristics
(e.g. contents in the same YouTube  category) \cite{Crane}.
Hence, restricting the analysis to a single class $m$ of contents,
we may assume that: $i)$ $Z_m$ represents the demand volume, i.e. the total number of requests it typically attracts;
$ii)$ all contents of the class exhibit the same normalized popularity profile. This justifies the choice of a SNM
with a multiplicative popularity profile such as \eqref{eq:popprof}.

Recall that for this model the function $g$ is given by \eqref{eqg}.
Assuming \eqref{eq:hintthintmeanz}, by \eqref{eq:pinche} and \eqref{phitche} (with $\theta$ in place of $t_C$)
we easily have
\begin{align}\label{che_SNM-mm}
p_{\mathrm{hit},\mathrm{Che}}(\theta)=1-(\EE[Z_1])^{-1}\int_{0}^\infty h(u) \phi_{Z_1}'\left(-\int_{u-\theta}^{u}h(s)\,\mathrm{d}s \right) \mathrm{d} u,
\end{align}
where $\phi_{Z_1}'$ is the first order derivative of $\phi_{Z_1}$.
Relations \eqref{eqg} and \eqref{che_SNM-mm} provide a computationally efficient tool to estimate
the ``hit" probability, under Che's approximation, of LRU caches under the SNM. Indeed, we may estimate $t_C$ by numerically inverting \eqref{eqg} and using the relation $t_C= g^{-1}(C/\lambda)$.
Replacing $\theta$ in \eqref{che_SNM-mm} with such estimate of $t_C$, we finally have an estimate of the ``hit" probability under Che's approximation.

We now assess the accuracy of the Che approximation
for the  evaluation of the ``hit" probability by describing some numerical results.
We suppose that the arrival rate of new contents $\lambda$ is
equal to $100.000$ units per day; 
we assume that the demand volume $Z_1$
follows a Pareto distribution with probability density $f_{Z_1}(z)=\alpha a^\alpha/z^{1+\alpha},\quad z\geq a>0,\,\alpha>1,$
and mean
$\mathbb{E}[Z_1]=\frac{\alpha a}{\alpha-1}=3$ (we refer the reader to \cite{Roberts12} and \cite{nostro-tec-rep} for a practical justification
on the choice of a Pareto distribution); we consider a multiplicative popularity profile
of the form \eqref{eq:popprof} with $h(t):=\frac{1}{L}\ind\{0 \le t\leq L\}$,
where the parameter $L$ has to be interpreted as the content
life-span.

\tgifeps{10}{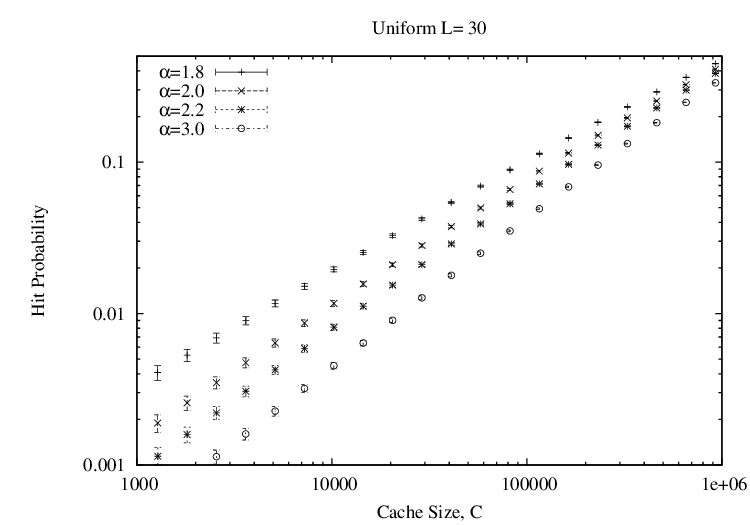}{$p_{hit}$ vs cache size for different values of the exponent $\alpha>1$ and content life-span $L=30$.}
\tgifeps{10}{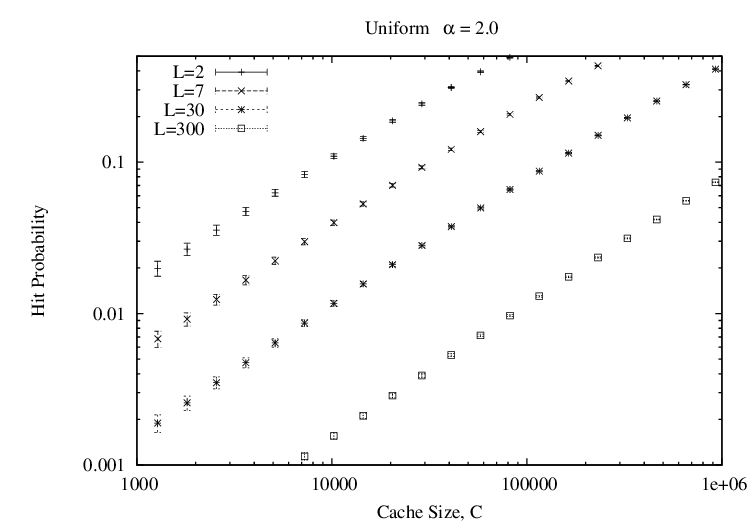}{$p_{hit}$ vs cache size for different values of the content life-span $L$ and exponent $\alpha=2$.}

\begin{table}
\begin{tabular}{|cr|cccc|cccc|}
\hline
   $\alpha $ & $L$  & $C$ & $p_{\text{hit,Che}}$ &  $\underline{p_{\text{hit}}}$  & $\overline{p_{\text{hit}}}$ & C   &    $p_{\text{hit,Che}}$ &  $\underline{p_{\text{hit}}}$  & $\overline{p_{\text{hit}}}$\\
 \hline
 \hline

 $1.8$ & 30 &10240 & 0.019596 &	 0.018880 &	  0.020313 & 163840 & 0.144328 	& 0.143126 &	  0.145529 \\
 $2.0$ & 2 &10240  & 0.109252 &	 0.105353 &	  0.113151 &   163840 &  0.671657 	& 0.669498 &	  0.673815   \\
 $2.0$ & 7 &10240  &       0.039790 &	 0.038232	&  0.041348   & 163840& 0.343061 	& 0.340319	&  0.345802  \\
  $2.0$ & 30 &10240  & 0.011657 &	 0.011158	&  0.012156  & 163840  & 0.114597 &	 0.113516 &	  0.115677\\

  $2.0$ & 300 &10240 & 0.001555 	& 0.001480	&  0.001629 & 163840 & 	0.017497 &	 0.017305&	  0.017688\\
$2.2$ & 30 &10240  & 0.008125 &	 0.007747 &  0.008504 &  163840 & 0.096641 &	 0.095651	&  0.097630 \\
$3.0$  & 30 & 10240 & 0.004524 &	 0.004293	&  0.004755 &   163840 &  0.068667 &	 0.067871	 & 0.069464 \\
\hline
\end{tabular}
\caption{Numerical values for the Che approximation  of the \lq \lq hit''  probability ($p_{\text{hit,Che}}$),
and for the lower ($\underline{p_{\text{hit}}}$) and the upper ($\overline{p_{\text{hit}}}$) bounds of the true \lq \lq hit'' probability.}
\label{Table1}
\end{table}

Figures \ref{fig:figura1new} and \ref{fig:figura2new} report the ``hit" probability, as predicted by Che's approximation,
vs the cache size for different values of the exponent $\alpha$ and the content life-span $L$, respectively.
For each estimate,
the figures show also the interval in which the exact value of the ``hit" probability falls as given by Proposition
\ref{prop:error0}.
All computations have been carried out   while guaranteeing  relative numerical errors smaller than $10^{-2}$.
Some selected results are additionally reported in Table \ref{Table1}.
Note that in all cases of practical relevance (i.e. for values of the ``hit" probability exceeding $10^{-2}$)
Che's approximation leads to negligible errors. The surprisingly good degree of accuracy entailed by Che's approximation, which has been
already experimentally (i.e. against simulations) observed by several authors \cite{Roberts12,nostro_infocom}, is now confirmed even for the SNM.

Further  numerical results providing  useful insights on the cache performance
can be found in \cite{nostro-tec-rep}.


\begin{thebibliography}{10}
\scriptsize

\renewcommand{\baselinestretch}{1.2}

\bibitem{bianchi}
G. Bianchi et al. 
\newblock Check before storing: what is the performance price of content integrity verification in LRU caching?
\newblock {\em ACM Comput. Comm. Rev.} 43, 59--67, 2013.

\bibitem{bondesson}
L. Bondesson.
\newblock Shot noise processes and shot noise distributions.
\newblock In: {\em Encyclopedia of Statistical Sciences} 8, 448--452, Wiley, New York.

\bibitem{bordenavetorrisi}
C. Bordenave and G.L. Torrisi.
\newblock Monte Carlo methods for sensitivity analysis of Poisson driven stochastic systems, and applications.
\newblock {\em Adv. Appl. Probab.} 40, 293-320, 2008.

\bibitem{che}
H. Che, Y. Tung and Z. Wang.
\newblock Hierarchical Web caching systems: modeling, design and experimental results.
\newblock {\em IEEE JSAC} 20, 1305--1314, 2002.

\bibitem{Coffman:73}
E. Coffman and P. Denning.
\newblock {\em Operating Systems Theory}.
\newblock Englewood Cliffs, New Jersey, 1974.

\bibitem{Crane}
R. Crane and D. Sornette.
\newblock Robust dynamic classes revealed by measuring the response function of a social system.
\newblock {\em PNAS} 105, 15649-15653, 2008.

\bibitem{daley}
D.J. Daley and D. Vere-Jones.
\newblock {\em An Introduction to the Theory of Point Processes. Vol. I}.
\newblock Springer, New York, 2003.

\bibitem{daley2}
D.J. Daley and D. Vere-Jones.
\newblock {\em An Introduction to the Theory of Point Processes. Vol. II}.
\newblock Springer, New York, 2008.

\bibitem{Dan1990}
A. Dan and D. Towsley.
\newblock An approximate analysis of the LRU and FIFO buffer replacement schemes.
\newblock {\em SIGMETRICS Perform. Eval. Rev.} 18, 143--152, 1990.

\bibitem{dembo}
A. Dembo and O. Zeitouni.
\newblock {\em Large Deviation Techniques and Applications}.
\newblock Springer, New York, 1998.

\bibitem{chunking}
P.M. Deshpande et al.
\newblock Caching multidimensional queries using chunks.
\newblock {\em ACM SIGMOD Record.} 27, 1998.

\bibitem{duffield}
N.G. Duffield and W. Whitt.
\newblock Large deviations of inverse processes with nonlinear scalings.
\newblock {\em Ann. Appl. Probab.} 8, 995--1026, 1998.

\bibitem{duffytorrisi}
K. Duffy and G.L. Torrisi.
\newblock Sample path large deviations of Poisson shot noise with heavy tail semi-exponential distributions.
\newblock{\em J. Appl. Probab.} 48, 688--698, 2011.

\bibitem{towsley-nain}
N.C. Fofack et al.
\newblock Performance evaluation of hierarchical TTL-based cache networks.
\newblock {\em Computer Networks} 65, 212-231, 2014.

\bibitem{Roberts12}
C. Fricker, P. Robert and J. Roberts.
\newblock A versatile and accurate approximation for LRU cache performance.
\newblock {\em Proceedings of ITC}, Krakow, Poland, pp. 1-8, 2012.

\bibitem{ganeshmaccitorrisi1}
A. Ganesh, C. Macci and G.L. Torrisi.
\newblock Sample path large deviations principles for Poisson shot noise processes, and
applications.
\newblock {\em Electron. J. Probab.} 10, 1026--1043, 2005.

\bibitem{ganeshtorrisi}
A. Ganesh and  G.L. Torrisi.
\newblock A class of risk processes with delayed claims: ruin probability estimates under heavy tail conditions.
\newblock {\em J. Appl. Probab.} 43, 916-926, 2006.

\bibitem{ganeshmaccitorrisi2}
A. Ganesh, C. Macci and G.L. Torrisi.
\newblock A class of risk processes with reserve-dependent premium rate: sample path large deviations and importance sampling.
\newblock{\em Queueing Systems} 55, 83-94, 2007.

\bibitem{Jacobson-ICN}
V. Jacobson et al.
\newblock Networking named content.
\newblock {\em ACM CoNEXT}, Rome, Italy, 2009.

\bibitem{Jiang_conext12}
W. Jiang et al., 
\newblock Orchestrating massively distributed CDNs.
\newblock {\em ACM CoNEXT}, Nice, France, 2012.

\bibitem{virtamo98}
K. Kylakoski and J. Virtamo.
\newblock Cache replacement algorithms for the renewal arrival model.
\newblock {\em Proc. of Fourteenth Nordic Teletraffic Seminar}, Copenhagen, Denmark, pp. 139-148, 1998.

\bibitem{nostro-tec-rep}
E. Leonardi and G.L. Torrisi.
\newblock Least Recently Used caches under the Shot Noise Model.
\newblock{\em IEEE INFOCOM}, Hong-Kong, China, pp. 2281-2289, 2015.

\bibitem{lowen}
S.B. Lowen and M.C. Teich.
\newblock Power-law shot noise.
\newblock {\em IEEE Trans. Inform. Theory} 36, 1302--1318, 1990.

\bibitem{nostro_infocom}
V. Martina et al. 
\newblock A unified approach to the performance analysis of caching systems.
\newblock {\em IEEE INFOCOM}, Toronto, Canada, pp. 2040-2048, 2014.

\bibitem{mollertorrisi}
J. Moller and G.L. Torrisi.
\newblock Generalised shot noise Cox processes.
\newblock{\em Adv. Appl. Probab.} 37, 47-74, 2005.

\bibitem{nourdin}
I. Nourdin and G. Peccati.
\newblock {\em Normal approximation with Malliavin calculus}.
\newblock Cambridge University Press, Cambridge, 2012.

\bibitem{peccati}
G. Peccati, J.L. Sol{\'e}, M.S. Taqqu and F. Utzet.
\newblock Stein's method and normal approximation of Poisson functionals.
\newblock {\em Ann. Probab.} 38, 443--478, 2010.

\bibitem{penrose}
M. Penrose.
\newblock {\em Random geometric graphs}.
\newblock Oxford University Press, New York, 2004.

\bibitem{ICN-1}
I. Psaras, W.K. Chai and G. Pavlou.
\newblock Probabilistic methods in network caching for information-centric networks.
\newblock {\em ICN workshop on Information-centric networking}, 2012.

\bibitem{kurose}
E.J. Rosensweig et al.
\newblock Approximate models for general cache networks.
\newblock {\em IEEE INFOCOM}, San Diego, USA, pp. 1-9, 2010.

\bibitem{stabiletorrisi}
G. Stabile and G.L. Torrisi.
\newblock Large deviations of Poisson shot noise processes, under heavy tail semi-exponential conditions.
\newblock{\em Statist. Probab. Let.} 80, 1200-1209, 2010.

\bibitem{torrisi}
G.L. Torrisi.
\newblock Simulating the ruin probability of risk processes
with delay in claim settlement.
\newblock {\em Stoch. Proc. Appl.} 112, 225--244, 2004.

\bibitem{nostroCCR}
S.Traverso et al. 
\newblock Temporal locality in today's content caching: why it matters and how to model it.
\newblock {\em  ACM Comput. Comm. Rev.} 43, 5--12, 2013.

\bibitem{nostro-TMM}
S. Traverso et al.
\newblock Unravelling the impact of temporal and geographical locality in content caching systems.
\newblock {\em IEEE Transactions on Multimedia} 17, 1839-1854, 2015.
\end{thebibliography}
\end{document}